\documentclass[runningheads]{llncs}
\usepackage{graphicx}
\usepackage{tabularx}
\usepackage[
  separate-uncertainty = true,
  multi-part-units = repeat
]{siunitx}
\usepackage{amssymb}
\usepackage{array}
\usepackage{microtype}

\begin{document}
\title{Automated fetal brain extraction from clinical Ultrasound volumes using 3D Convolutional Neural Networks}
\author{Felipe Moser\inst{1} \and
Ruobing Huang\inst{1} \and
Aris T. Papageorghiou\inst{2}\and
Bart\l{}omiej W. Papie\.{z} \inst{1, 3} \and
Ana I. L. Namburete\inst{1}}
%\author{ }
%\institute{ }
\authorrunning{ }
\titlerunning{ }
\institute{Institute of Biomedical Engineering, Department of Engineering Science, University of Oxford, Oxford, UK \and 
Nuffield Department of Obstetrics and Gynaecology, John Radcliffe Hospital, University of Oxford, Oxford, UK \and
Big Data Institute, Li Ka Shing Centre for Health Information and Discovery, University of Oxford, Oxford, UK}
\maketitle              % typeset the header of the contribution
\begin{abstract}
To improve the performance of most neuroimiage analysis pipelines, brain extraction is used as a fundamental first step in the image processing. But in the case of fetal brain development, there is a need for a reliable US-specific tool. In this work we propose a fully automated 3D CNN approach to fetal brain extraction from 3D US clinical volumes with minimal preprocessing. Our method accurately and reliably extracts the brain regardless of the large data variation inherent in this imaging modality. It also performs consistently throughout a gestational age range between 14 and 31 weeks, regardless of the pose variation of the subject, the scale, and even partial feature-obstruction in the image, outperforming all current alternatives.

\end{abstract}
\section{Introduction} \label{sec:introduction}
The use of Ultrasound (US) to study fetal brain development and to diagnose central nervous system malformations has been around for decades, and has become standard clinical practice around the world thanks to its ability to capture the development of the brain in the womb~\cite{Kim2008}~\cite{Haratz2018}~\cite{Namburete2018-2}. 3D US expands on this technique by allowing for the imaging of the whole brain at once, instead of one slice at a time. However, the positional variation of the brain inside the scan volume, as well as the large amount of extra-cranial tissue observed in the volume, constitute a serious challenge when analysing the data. This has led brain-extraction tools to become a fundamental first step in most neuroimage analysis pipelines with several methods being developed for fetal Magnetic Resonance Imaging (MRI) data but a reliable US-specific tool has not yet been developed. 

Another challenge is that during gestation, the fetus is constantly moving and rotating. This causes the position and orientation of the brain to vary drastically from measurement to measurement. In order to compensate for this high degree of variability, the standard clinical protocol is to position the sagittal plane in the middle of the 3D US volume and align the axial and coronal planes to the remaining axes.  While this reduces the variability to a degree, it is not consistent  as it depends entirely on the clinician to be accurate. There is therefore a need for a method that can accurately and reliably determine the position, orientation, and volume of the brain from a 3D US scan.

Besides the variation in position of both the fetus and the probe, the development of the brain throughout gestation increases the variability of the data, since the scale, ossification, and structural characteristics inside the skull change for each gestational week. The physical interaction of the US beam with the increasingly ossifying skull also causes reverberation artefacts and occlusions. Most importantly, when imaging from one side of the skull, the brain hemisphere farthest from the probe (distal) is visible while the closest (proximal) hemisphere is mostly occluded~\cite{ISUOG2007}. 

Several methods have been developed for the purpose of brain extraction (a comprehensive comparison of them for neonatal brain extraction can be found in Serag et al.~\cite{Serag2016}), but very limited amount of work has been done in relation to the extraction of the brain volume from fetal imaging and the vast majority of it has been focused on MRI imaging. Publications such as~\cite{Ison2012} and~\cite{Keraudren2013} show the difficulty
in developing a reliable method that can accurately locate the brain from the acquired images. To our knowledge, the only method developed for automated fetal brain extraction from 3D Ultrasound is the one proposed by Namburete et al., 2018~\cite{Namburete2018}. This method uses a fully-convolutional neural network (FCN) to predict the position of the brain from 2D slices of the 3D US volume. This prediction is then used to generate a 3D probability mask of fetal brain localization. An ellipsoid is then registered to the mask, resulting in an approximation of the location, orientation, and spatial extent of the brain. 
While this method offers a solution to the problem of brain extraction, it still relies on a 2D approach and its ellipsoid approximation does not accurately represent the shape of the brain.

Here, we propose an end-to-end 3D Convolutional Neural Network (CNN) approach for automated brain localization and extraction from standard clinical 3D fetal neurosonograms. As opposed to Namburete et al.~\cite{Namburete2018}, this method is a fully 3D approach to brain extraction and requires minimal pre- or post-processing.
We show that our network manages to accurately detect the complete brain with a high degree of accuracy and reliability, regardless of the gestational age (ranging from 14 to 30 weeks), brain and probe positions, and visible brain structures.

\section{Brain Extraction Network}
\begin{figure}[htb]
\centering
  \begin{tabular}{@{}c@{}}
    \includegraphics[width=\textwidth]{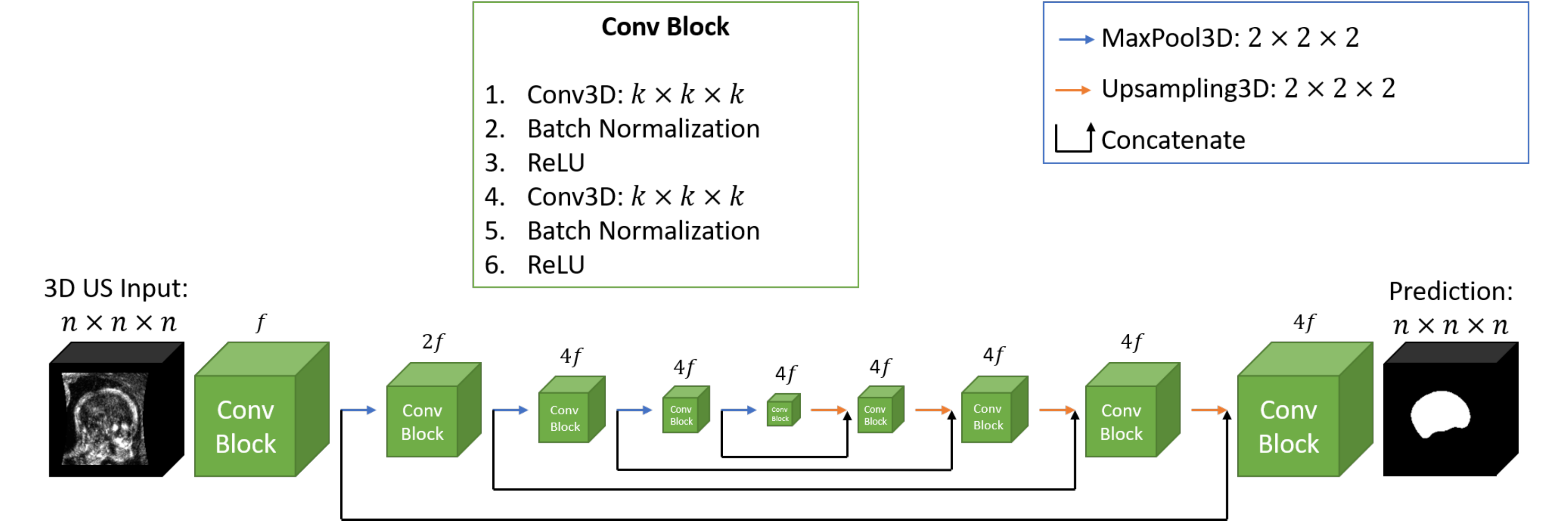}
\end{tabular}
  \caption{Schematics of the 3D CNN used for brain extraction. Each convolution block performes the steps described, with the numbers of filters used being a multiple of the first bock's filters $f$ and displayed above each block. This particular example shows four pooling layers (MaxPooling and UpSampling).  \label{fig:cnnschematics}}
\end{figure}

The general schematics of the CNN used for this work can be seen in Fig. \ref{fig:cnnschematics}. It is a variation of the network used in Huang et al., 2018~\cite{Huang2018} and is similar in structure to the 3D U-Net network from~\cite{Cicek2016}. This network design showed accurate and stable results for extracting brain-structures from 3D US scans and was therefore chosen as the starting point for this work. Our network comprises a kernel of size $k^3$, $l$ convolutional and down-sampling layers, and $l$ convolutional and up-sampling layers. The first two convolutional layers have $f$ and $2f$ number of filters, with the remaining ones having $4f$. After each convolution, batch normalisation and ReLu activationes were used. The network was trained end-to-end with the Adam optimizer, with a learning rate of $10^{-3}$ and a Dice-Loss function. Both input and output are of size $n \times n \times n$. 

\section{Experiments}

\subsection{Data}\label{sec:data}
\begin{figure}[htb]
\centering
  \begin{tabular}{@{}c@{}}
    \includegraphics[width=0.7\textwidth]{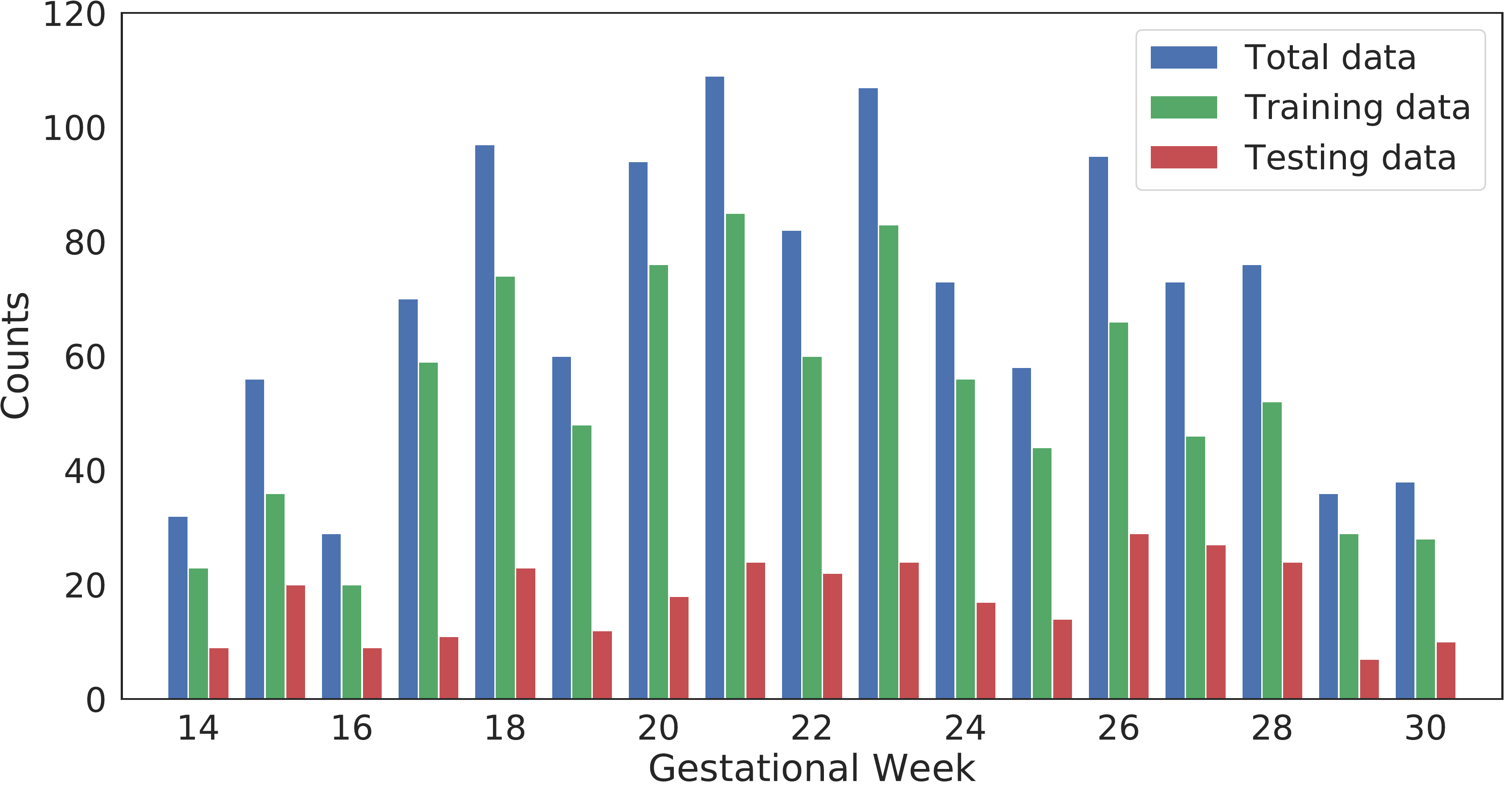}
\end{tabular}
  \caption{Histogram of the number of volumes for each gestational week. \label{fig:agehistogram}}
\end{figure}

A total of N = 1185 fetal brain volumes spanning between 14.4 and 30.9 weeks of gestation were used for the training and testing of our networks. The distribution of the gestational ages is shown in Fig. \ref{fig:agehistogram}. These volumes were obtained from INTERGROWTH-21st~\cite{Papageorghiou2014} and are from healthy subjects that experienced a normal pregnancy.
The original volumes have a median size of 237x214x174 voxels. They were centre-cropped to a size of 160x160x160 voxels and resampled to an isotropic voxel size of 0.6 mm x 0.6 mm x 0.6 mm.

In order to train our networks, labelled data representing the location of the brain within the 3D US volume were required. Manual labelling by a clinician would be time-consuming and is likely to have a large degree of uncertainty due to the artefacts and occlusions that are intrinsic to this imaging modality.

To circumvent this problem, the spatiotemporal fetal brain atlases generated by Gholipour et al., 2017~\cite{Gholipour2017} were used as templates. These atlases were generated from fetal MRI data. It represents the average brain shape for each gestational week and allows us to use it for semi-automatic annotations.

To annotate our data, the complete dataset was aligned by hand and rigidly scaled to the same coordinate system. Each brain mask from the atlas was isotropically scaled to a selected reference using a similarity transformation to preserver aspect ratio. The mask was then transformed back to the original position of the brain in the 3D US scan by performing the inverse transformation that was used to align the volumes in the first place. This resulted in a consistent brain annotation for the complete dataset.
Since the Gholipour's atlases span a gestational age range of 21 to 38 weeks, the atlas of week 21 was used to represent gestational weeks 14 to 18 of our dataset.

\subsection{Performance evaluation}\label{sec:perfeval}
To determine the best network for fetal brain extraction, eight networks with different hyperparameters were created: Input size $n$ (same in all dimensions), number of pooling layers $l$, kernel size $k$ (same in all dimensions), and number of filters of the first convolution $f$. For three-fold cross-validation experiments, 885 training volumes were partitioned into a subset of 685 for training and 200 for validation, three times. The best model was then retrained with the full 885 volumes and tested with the 300 hold-out testing set. A T-test's result of t = 1.26, p = 0.21 confirmed no statistically significant difference in their age distributions.

To evaluate the performance of each network, we used four separate measures that could give an overall assessment of reliability. Firstly, the Euclidean Distance (ED) between the centre of mass of the binary masks was calculated. This value gives a measure of the distance between the predicted centre and known centre of the brain. Since the position of the fetal brain relative to the 3D US volume varies throughout our dataset, the ED is a good indication of the accuracy of the predicted location of the brain. Secondly, the Hausdorff Distance (HD) between the masks was calculated. This metric determines the maximum distance between the prediction and the annotation and it allows to determine if there are areas that are mismatched, regardless of the accuracy of the rest of the predictions. Since the masks used for annotation come from fetal MRI data which contains different structural information than 3D US, the use of this metric is critical to assess the reliability of our network. Thirdly, the Dice Similarity Coefficient (DSC) was calculated to determine the amount of overlap between the annotation and the predicted binary mask (different threshold were explored). This coefficient gives a good general assessment of the accuracy of the prediction. And fourthly, a Symmetry Coefficient (SC) was calculated to represent the symmetry of the predcition. Since the interaction between the 3D US beam and the skull causes the proximal hemisphere to be occluded, the structural information within the skull is asymmetric. However, the skull is generally symmetric about the sagittal plane, and therefore a prediction that reflects this is imperative. To calculate the SC, the prediction masks were aligned to a common set of coordinates with the sagittal, axial, and coronal planes being the mid-planes of the volume (same used for the data annotation described in Sec. \ref{sec:data}). The right half of the brain prediction was then mirrored and the DSC between it and the left half is the value of the SC.

Using the fourth measures described (ED, HD, DSC, SC), we assessed the performance of our method under the intrinsic variability associated with fetal-brain 3D US. We first analysed the pose dependence of the accuracy of the prediction. As stated in Sec. \ref{sec:introduction}, one of the main challenges of automatically extracting the brain from a 3D US volume is the pose variation of both the fetus and the probe. To assess the reliability of our network in regards to the orientation of the fetal brain in the volume, the Euler angles needed for the alignment of each volume to a set of common coordinates was compared to the DSC of that particular prediction. This would determine if there is a correlation between the orientation and the quality of the prediction. 
We then determined the dependency of the network to the gestational week of the subject. Since the goal is to develop a model that can reliably extract the brain regardless of the gestation week of the subject, their accuracy of the prediction should be consistent throughout the 14.4 to 30.9 weeks of gestation that comprises our dataset. To assess this, the results from the first four experiments were divided into gestational weeks and compared.
Finally, we compared the predictions to the annotations by observing the regions of false-positives (predicted voxels not present in the annotation) and false-negatives (non-predicted voxels present in the annotations) results. This is a good qualitative method to analyse the regional accuracy of the prediction.

\section{Results}
\subsection{Cross-validation and network selection}\label{sec:crossval}
A description of each tested network, as well as the cross-validation results of each can be seen in Table \ref{tab:crossval}. 
The best score in terms of the ED between the centre of mass of the ground truth binary volume and the prediction of the network is from network D (1.16 mm), closely followed by network A. All networks managed to achieve a prediction with ED bellow 2 mm. For comparison, the median occipitofrontal diameter for 22 weeks of gestation (the mean gestational age of our dataset) is 62 mm~\cite{Moreira2011}.
The HD results show network D outperforming all other networks again with a result of 8.46 mm, followed by network F.
Finally, the DSC results are very consistent among all networks, managing to stay above 0.90, and achieving a maximum of 0.94 for networks A, D, E, and F. 
Network D obtained the best cross-validation results across all tests and was therefore selected to be trained with the full dataset. 

\begin{table}[h]
\caption{Description of the eight tested networks as well as the cross-validation results. Network D, in bold, shows the best results. $n$: Input size (same in all dimensions). $l$: Number of pooling layers. $k$: Kernel size (same in all dimensions). $f$: Number of filters of the first convolution.  ED: Euclidean Distance between centres of masses. HD: Hausdorff distance. DSC: Dice Similarity Coefficient. Param.: Number of trainable parameters in the network. \label{tab:crossval}}
\begin{tabular*}{\textwidth}{l@{\extracolsep{\fill}}llllllr}
  \hline
  Network ($n$)& $l$ & $k$ & $f$ & ED [mm]& HD [mm]& DSC & Param.\\
    \hline \\[-0.1in]
  A (80)& 4& 3 & 16& \SI{1.17 \pm 0.67}{}& \SI{10.19 \pm 6.40}{}& \SI{0.94 \pm 0.02}{}& 1.6 M\\
  B & 4 & 5 & 16& \SI{1.34 \pm 0.95}{}& \SI{9.53 \pm 4.37}{}& \SI{0.93 \pm 0.03}{}& 7.4 M\\
  C & 4 & 7 & 16& \SI{1.33 \pm 0.67}{}& \SI{9.75 \pm 4.30}{}& \SI{0.93 \pm 0.03}{}& 20.3 M\\
 \textbf{D} & \textbf{4} & \textbf{3 } & \textbf{8} & \textbf{1.16$\pm$0.69}& \textbf{8.46$\pm$3.66}& \textbf{0.94$\pm$0.02}& \textbf{0.4 M}\\
  E & 4 & 3 & 4& \SI{1.25 \pm 0.81}{}& \SI{9.82 \pm 6.01}{}& \SI{0.94 \pm 0.02}{}& 0.1 M\\[0.03in] \hline \\[-0.1in]
  F (160)& 4 & 3 & 4 &\SI{1.24 \pm 0.73}{}& \SI{9.14 \pm 4.53}{}& \SI{0.94 \pm 0.02}{}& 0.1 M\\
  G & 3 & 3 & 4& \SI{1.33 \pm 0.72}{}& \SI{12.12 \pm 6.87}{}& \SI{0.93 \pm 0.02}{}& 0.07 M\\
  H & 2 & 3 & 4& \SI{1.81 \pm 1.26}{}& \SI{23.12 \pm 6.99}{}& \SI{0.90 \pm 0.05}{}& 0.03 M\\
  \hline
\end{tabular*}
\end{table}

\subsection{Testing}
As shown in Sec. \ref{sec:crossval}, network D outperformed other networks, and so it was used for further experiments. It was trained with the full 885 training volumes and tested with the independent test set of 300 volumes. The testing results are shown in Table \ref{tab:testing}. Five different predictions threshold were tested to find the best results.

HD and SC values of threshold 1 are the best ones. However, this threshold is too high and makes the DSC fall when compared to the overall best threshold of 0.5. The latter, while having a slightly worse HD and SC, has a slightly better ED, and it has a statistically significant improvement of the DSC. Considering the minimal variation of ED and HD throughout the thresholds, using threshold 0.5 is the most appropriate, since it achieves a high degree of overlap (0.94) with the annotations, while keeping a good symmetry (0.95). 

Throughout the different thresholds, the results of our network are very consistent. This shows the high degree of confidence of the predictions generated. The results are also very consistent with the ones observed during cross-validation (see Table \ref{tab:crossval}), which confirms that the network works with new data.

\begin{table}[h]
\caption{Testing results of the fully-trained network. A threshold of 0.5 (in bold) showed the most consistent results and was therefore the best. ED: Euclidean Distance between centres of masses. HD: Hausdorff distance. DSC: Dice Similarity Coefficient. SC: Symmetry Coefficient \label{tab:testing}}
\begin{tabular*}{\textwidth}{l@{\extracolsep{\fill}}llll}
  \hline
Threshold & ED [mm]& HD [mm]& DSC & SC\\ \hline
	0 & \SI{8.34 \pm 3.26}{} & \SI{62.50 \pm 7.91}{} & \SI{0.27 \pm 0.15}{} & \SI{0.80 \pm 0.02}{}\\
	0.25 & \SI{1.43 \pm 0.93}{} & \SI{9.24 \pm 4.77}{} & \SI{0.94 \pm 0.02}{} & \SI{0.95 \pm 0.02}{}\\
	\textbf{0.5} & \textbf{1.36$\pm$0.72}& \textbf{9.05$\pm$3.56} & \textbf{0.94$\pm$0.02} & \textbf{0.95$\pm$0.02}\\
	0.75 & \SI{1.36 \pm 0.72}{} & \SI{8.97 \pm 3.54}{} & \SI{0.93 \pm 0.03}{} & \SI{0.80 \pm 0.02}{}\\
	1 & \SI{1.42 \pm 0.80}{} & \SI{8.72 \pm 4.23}{} & \SI{0.90 \pm 0.03}{} & \SI{0.99 \pm 0.01}{}\\

    \hline
\end{tabular*}
\end{table}

\subsection{Performance with pose variation}
\begin{figure}[!htb]
\centering
\scriptsize
  \begin{tabular}{@{}cccccc@{}}
    \includegraphics[width=.16\textwidth]{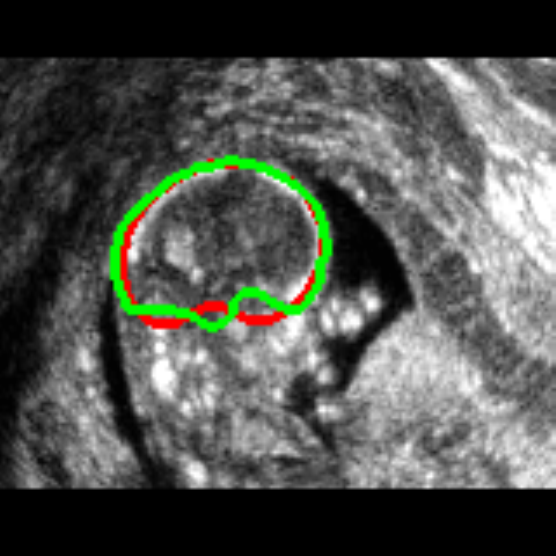} &
    \includegraphics[width=.16\textwidth]{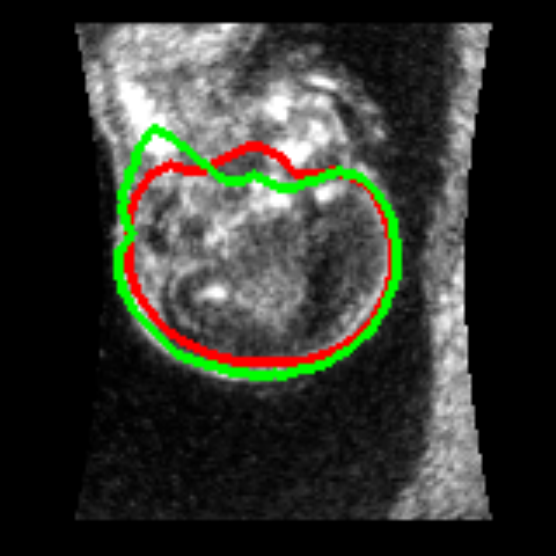} \vline &
    \includegraphics[width=.16\textwidth]{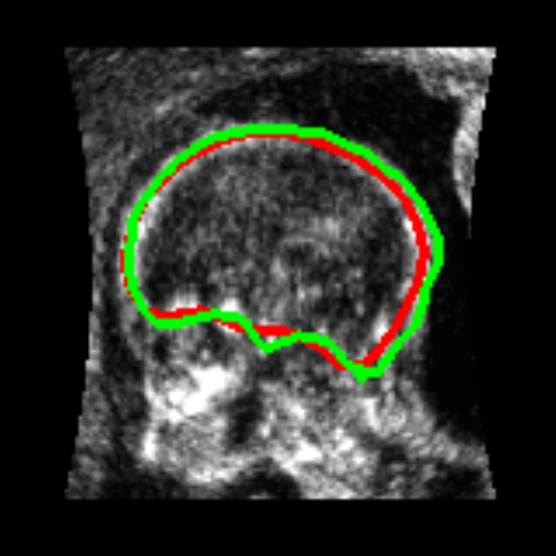} &
    \includegraphics[width=.16\textwidth]{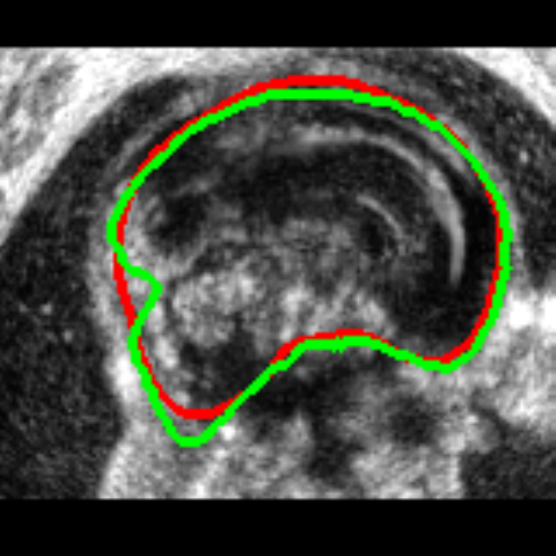} \vline &
    \includegraphics[width=.16\textwidth]{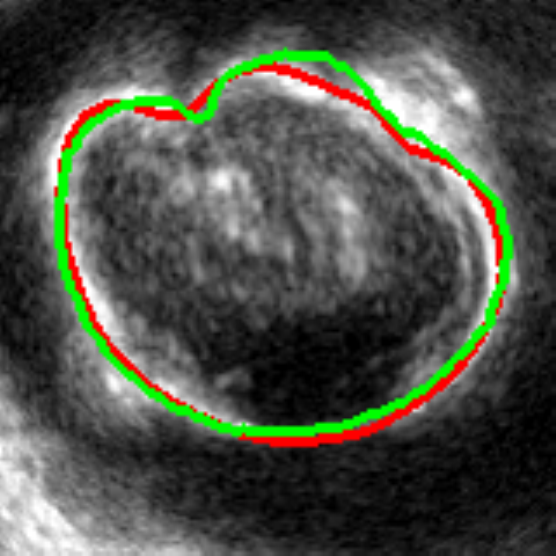} &
    \includegraphics[width=.16\textwidth]{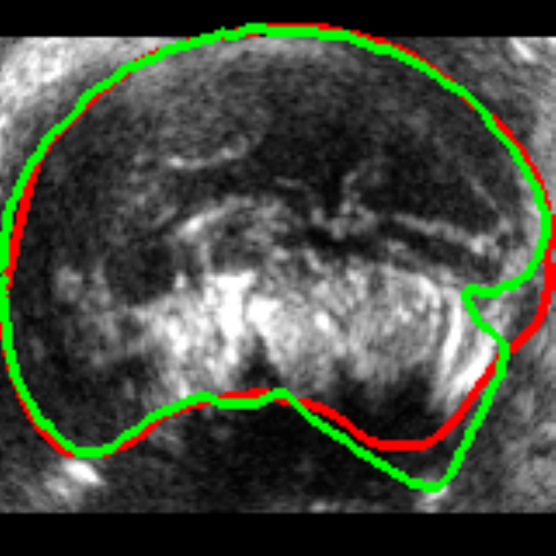} \\
    \includegraphics[width=.16\textwidth]{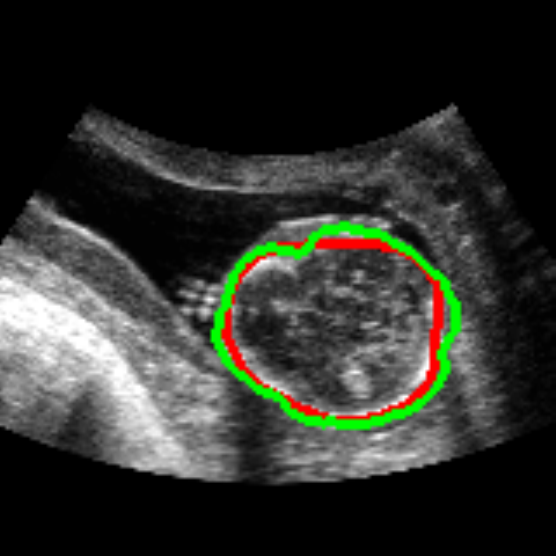} &
    \includegraphics[width=.16\textwidth]{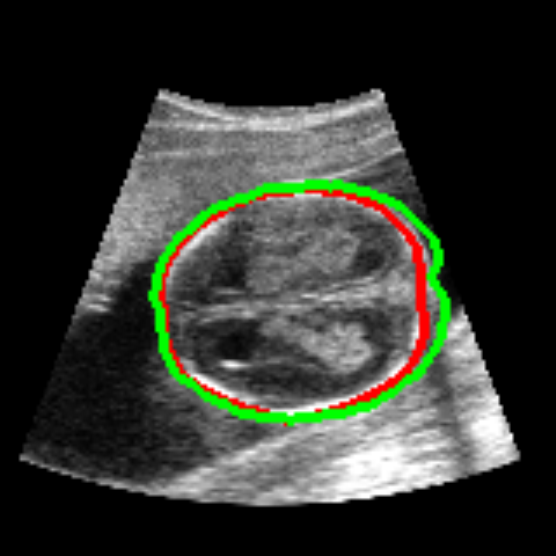} \vline &
    \includegraphics[width=.16\textwidth]{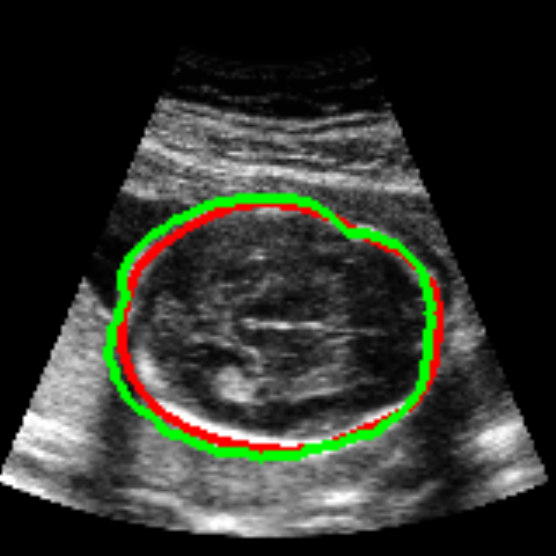} &
    \includegraphics[width=.16\textwidth]{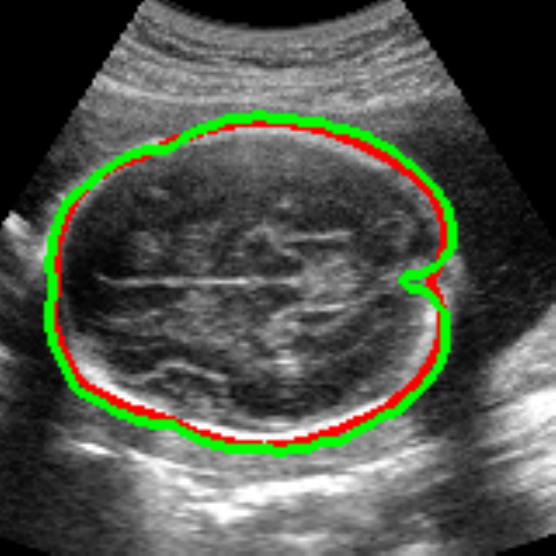} \vline &
    \includegraphics[width=.16\textwidth]{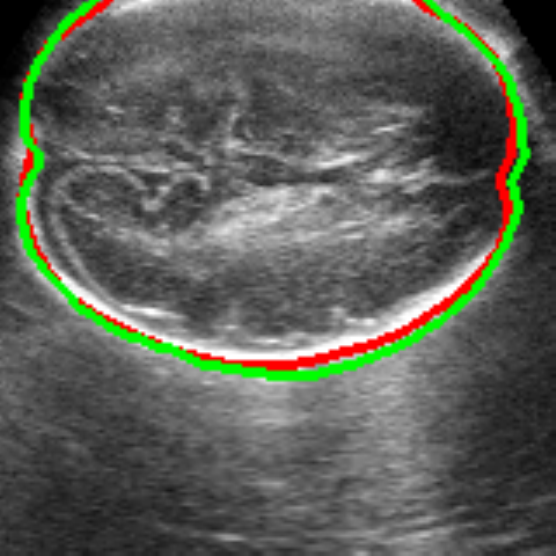} &
    \includegraphics[width=.16\textwidth]{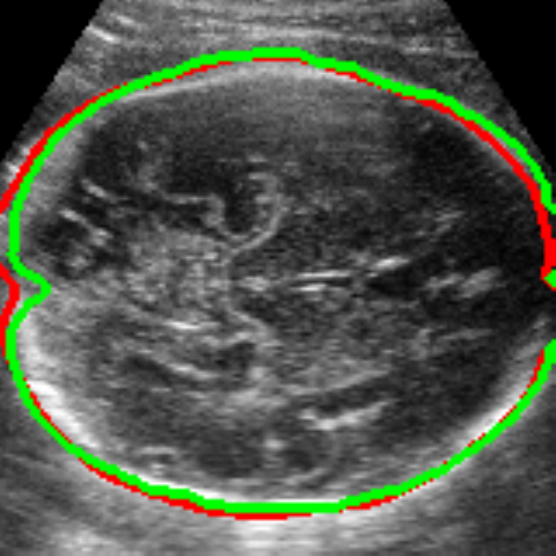}   \\
    \includegraphics[width=.16\textwidth]{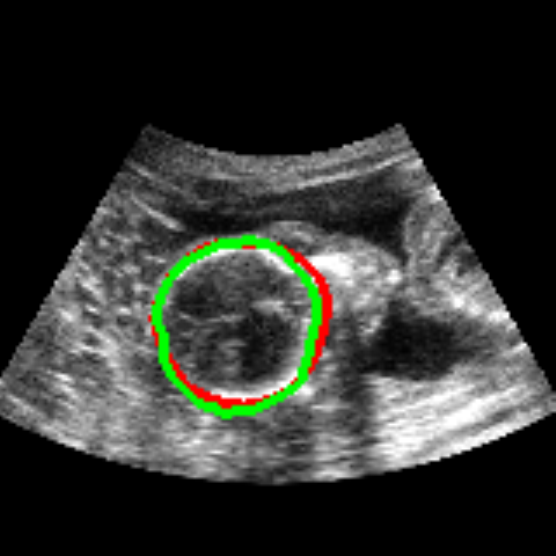} &
    \includegraphics[width=.16\textwidth]{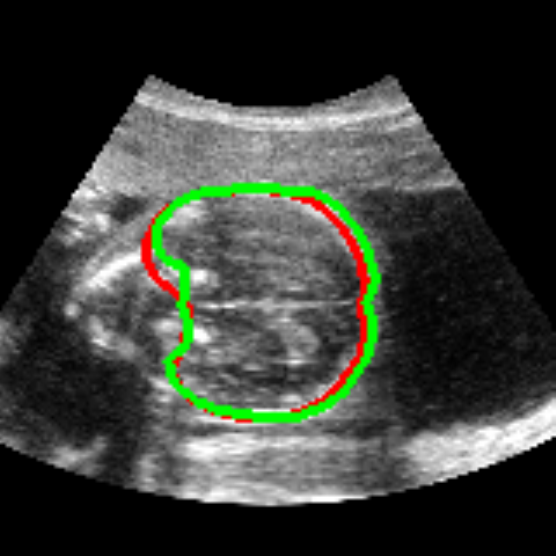} \vline &
    \includegraphics[width=.16\textwidth]{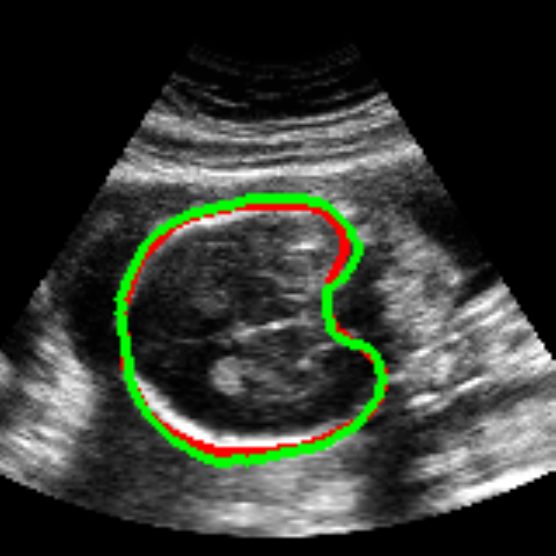} &
    \includegraphics[width=.16\textwidth]{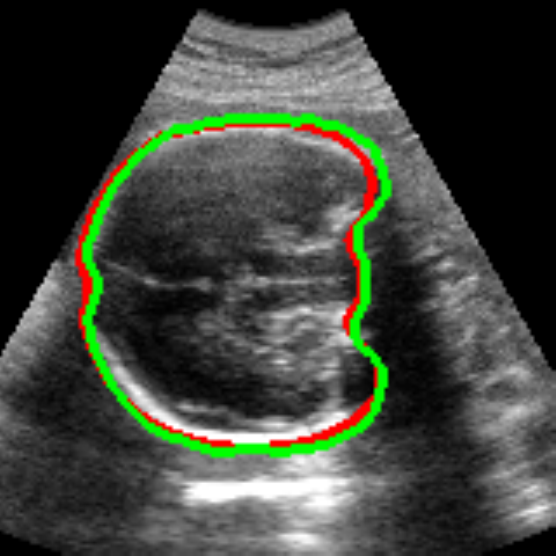} \vline&
    \includegraphics[width=.16\textwidth]{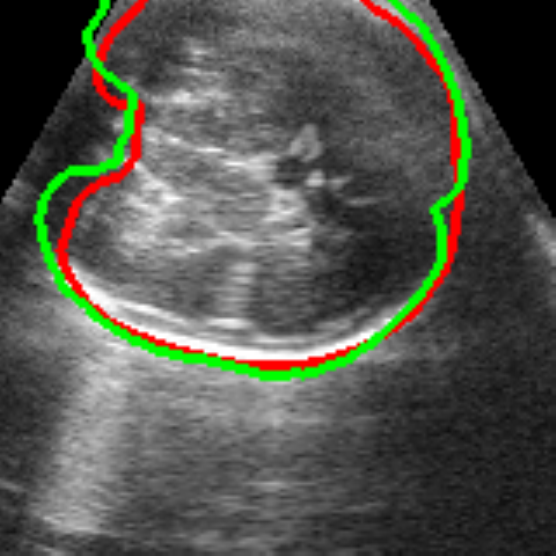} &
    \includegraphics[width=.16\textwidth]{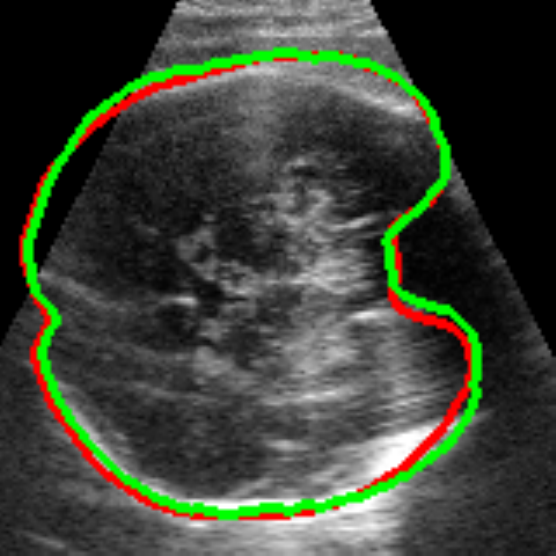}   \\
    \scriptsize{14 Weeks} & \scriptsize{17 Weeks} & \scriptsize{20 Weeks} & \scriptsize{22 Weeks} & \scriptsize{26 Weeks} & \scriptsize{30 Weeks}\\
    $\alpha$ = 53.8$^\circ$ &$\alpha$ = 84.4$^\circ$ &$\alpha$ = -64.6$^\circ$ &$\alpha$ = 82.3$^\circ$ &$\alpha$ = -61.4$^\circ$ &$\alpha$ = 75.5$^\circ$ \\
    $\beta$ = -68.0$^\circ$ &$\beta$ = 60.4$^\circ$ &$\beta$ = -60.6$^\circ$ &$\beta$ = -74.6$^\circ$ &$\beta$ = 59.0$^\circ$ &$\beta$ = -58.6$^\circ$ \\
    $\gamma$ = -39.0$^\circ$ &$\gamma$ = 92.2$^\circ$ &$\gamma$ = -114.2$^\circ$ &$\gamma$ = -83.6$^\circ$ &$\gamma$ = 119.6$^\circ$ &$\gamma$ = -107.8$^\circ$ \\
\end{tabular}
  \caption{Comparison of the brain-extraction prediction (red) and the ground-truth (green) superimposed onto the mid-planes of the 3D US volume. These volumes were selected to demonstrate the amount of variation between each scan. Top: XY-plane. Middle: XZ-plane. Bottom: YZ-plane. The gestation week of each volume and the Euler angles are displayed underneath. \label{fig:outlines}}
\end{figure}

As mentioned in Sec. \ref{sec:perfeval}, it is crucial for our brain extraction network to work consistently regardless of the orientation of the brain within the US volume. This can be qualitatively observed in Fig. \ref{fig:outlines}, which shows the outline of the brain-extraction prediction and the corresponding ground-truth, in red and green respectively, for six different 3D US volumes. These volumes have been selected to demonstrate the amount of variation between each scan, with the position of the fetus inside the mother as well as the position of the brain in respect to the scan varying drastically from case to case.

As shown in Fig. \ref{fig:outlines} , the network's prediction is remarkably close to the ground-truth, regardless of the position of the brain in the volume. It also manages to accurately predict the location of the brain when this is partially obscured either by the cropping or the shape of the ultrasound beam.

\begin{figure}[!htb]
\centering
  \begin{tabular}{cc}
    \includegraphics[height=0.4\textwidth]{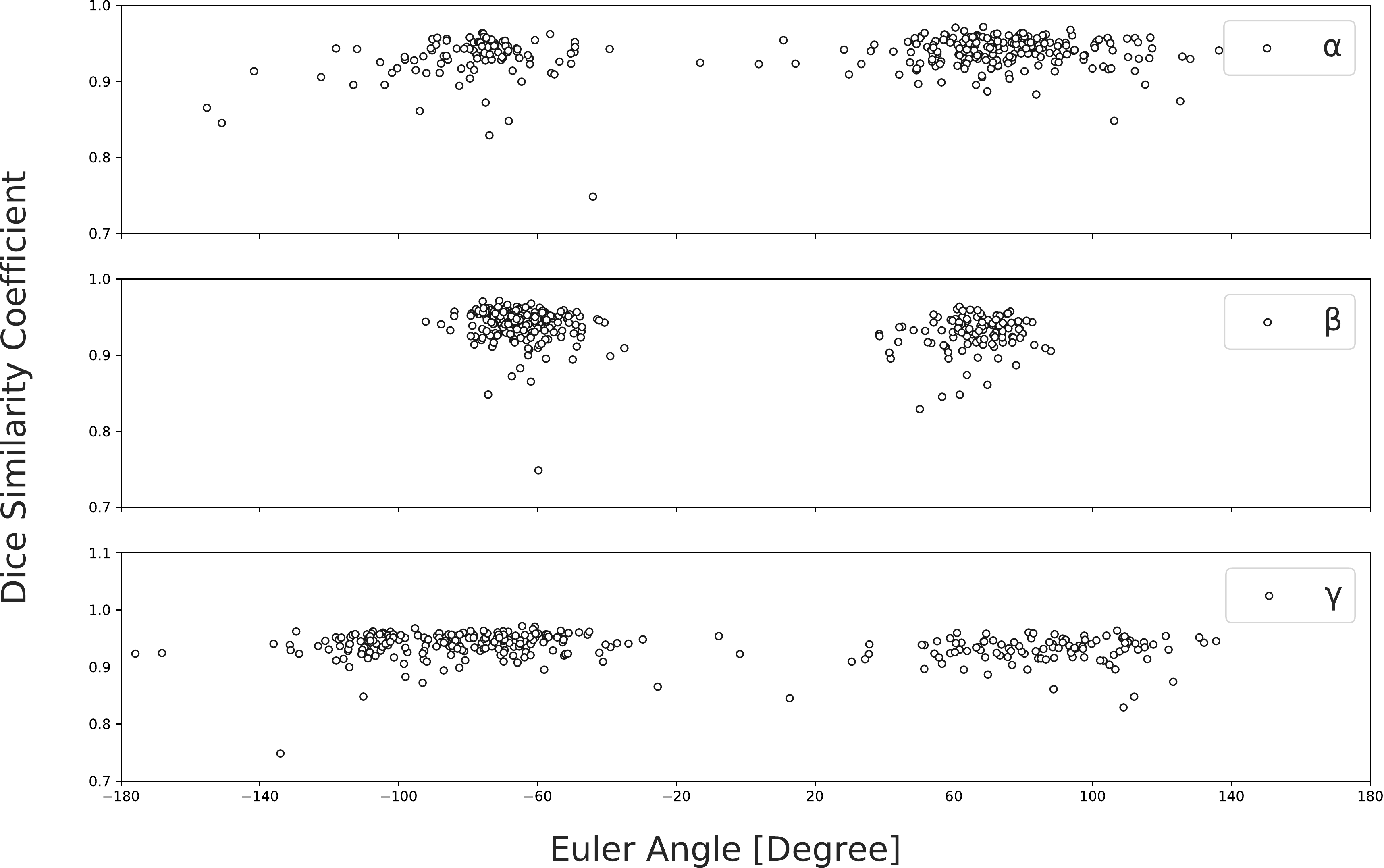} &
	\includegraphics[height=0.44\textwidth]{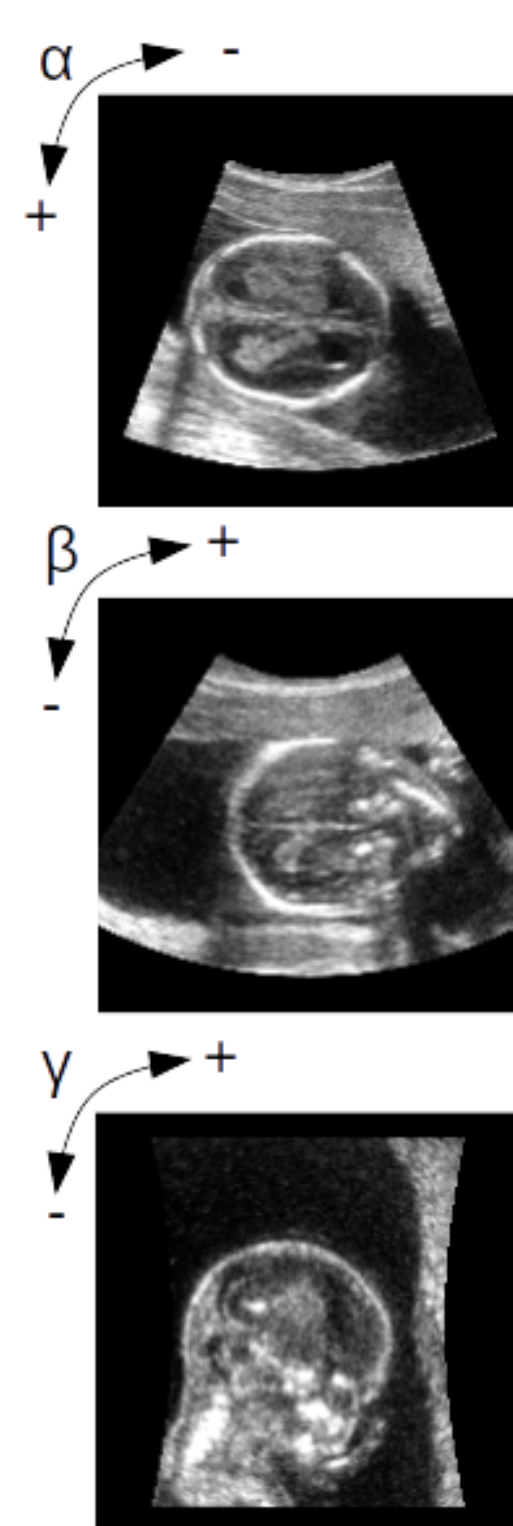}
\end{tabular}
  \caption{Plot of Euler angles vs DSC. The angles cluster around $90^\circ$ because of the manual alignment performed by the clinician being different to our coordinate system. No correlation between the Euler angles and the DSC was found. \label{fig:eulerdice}}
\end{figure}

The distribution of the DSC in relation to the Euler angles is shown in Fig. \ref{fig:eulerdice}. As expected, there is no visible correlation. The Pearson's correlation coefficient what calculated for each Euler angle, with a r$_\alpha$=0.15, r$_\beta$=-0.20, and r$_\gamma$=-0.16 respectively. The results show no significant correlation between the Euler angles and the DSC.


\subsection{Performance with different gestational ages}
\begin{figure}[!htb]
\centering
  \begin{tabular}{c@{\hskip 0.03\textwidth}c}
    \includegraphics[width=0.45\textwidth]{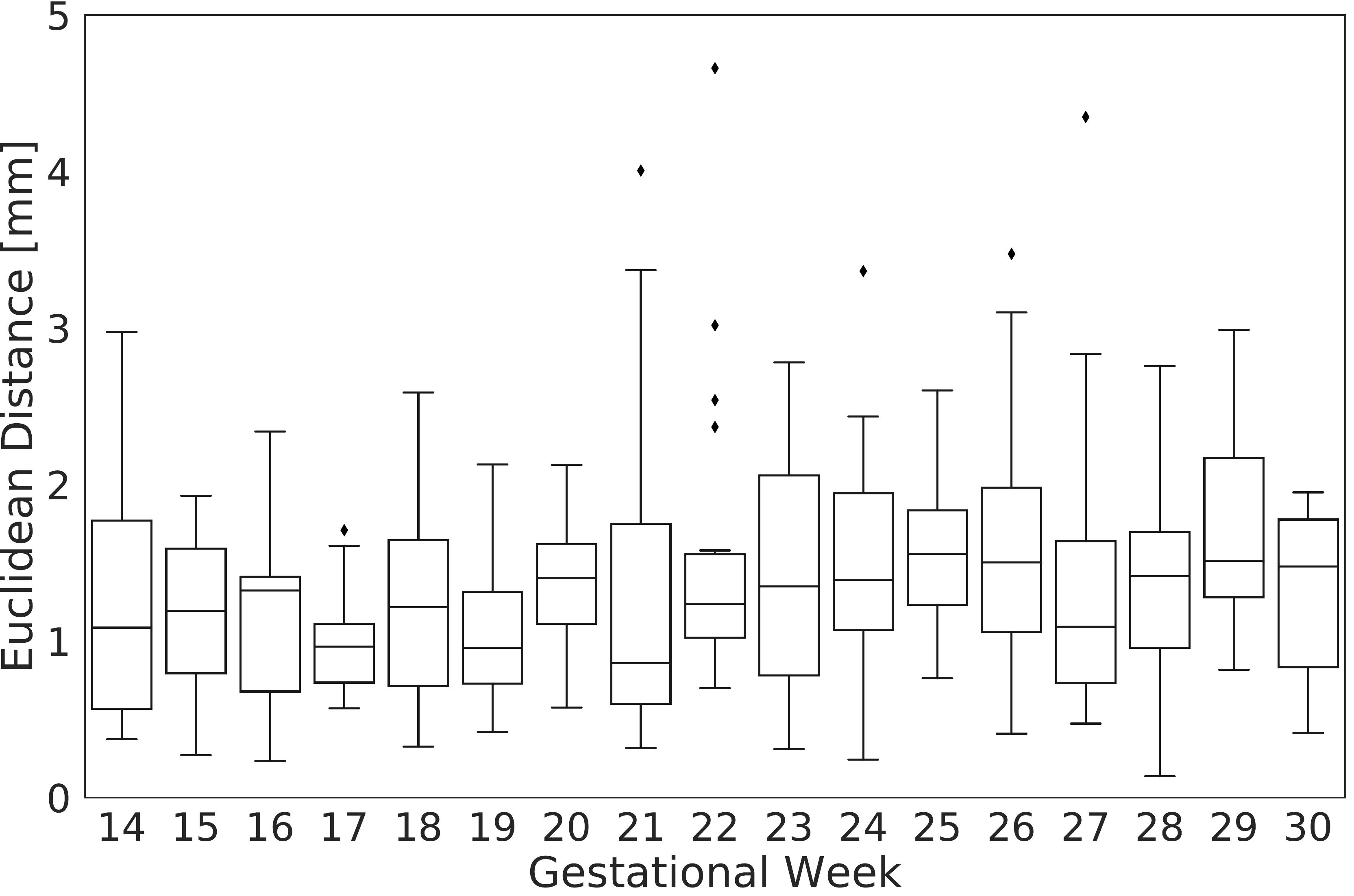} &
    \includegraphics[width=0.45\textwidth]{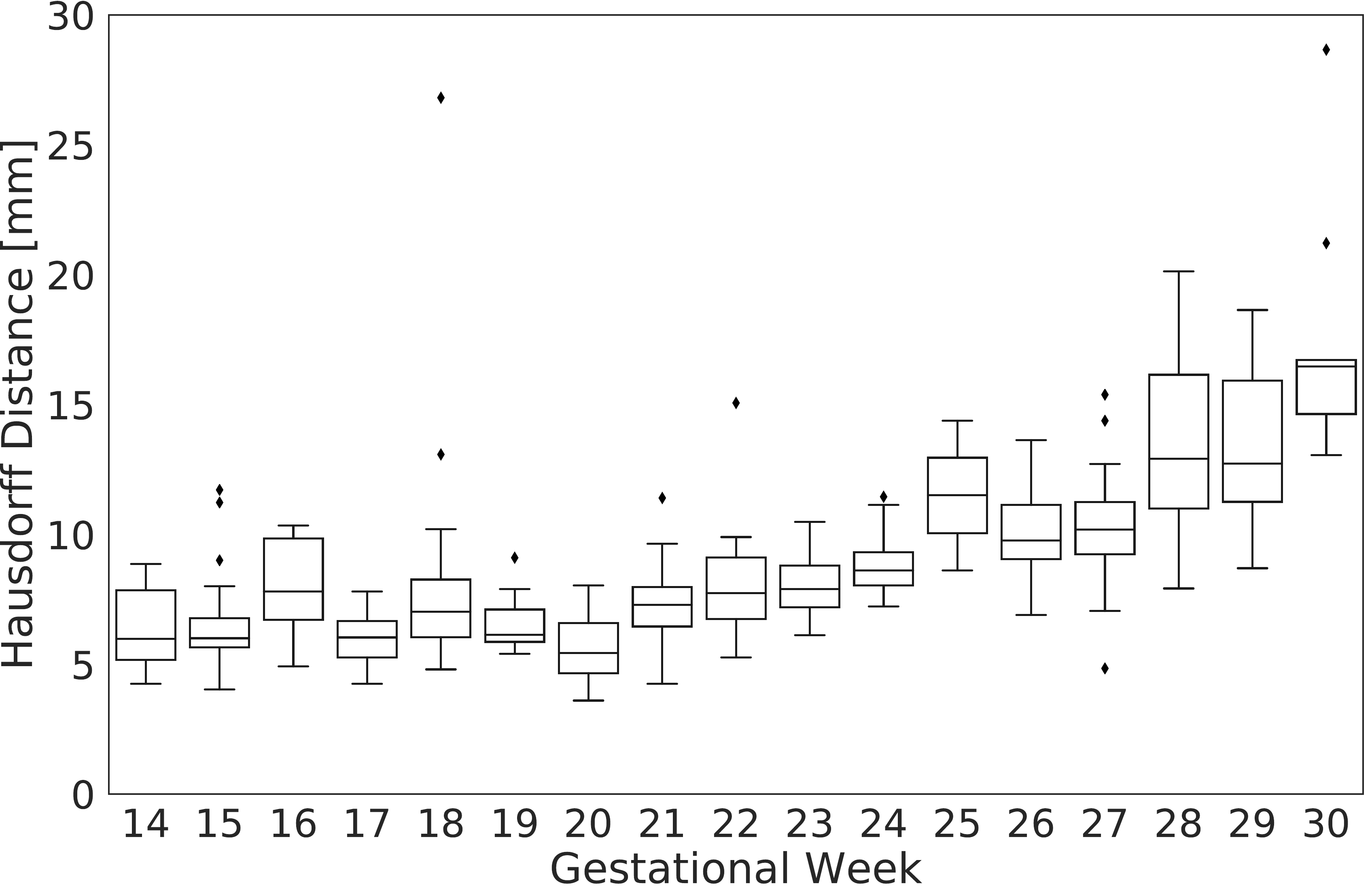} \\[0.02\textwidth]
    \includegraphics[width=0.45\textwidth]{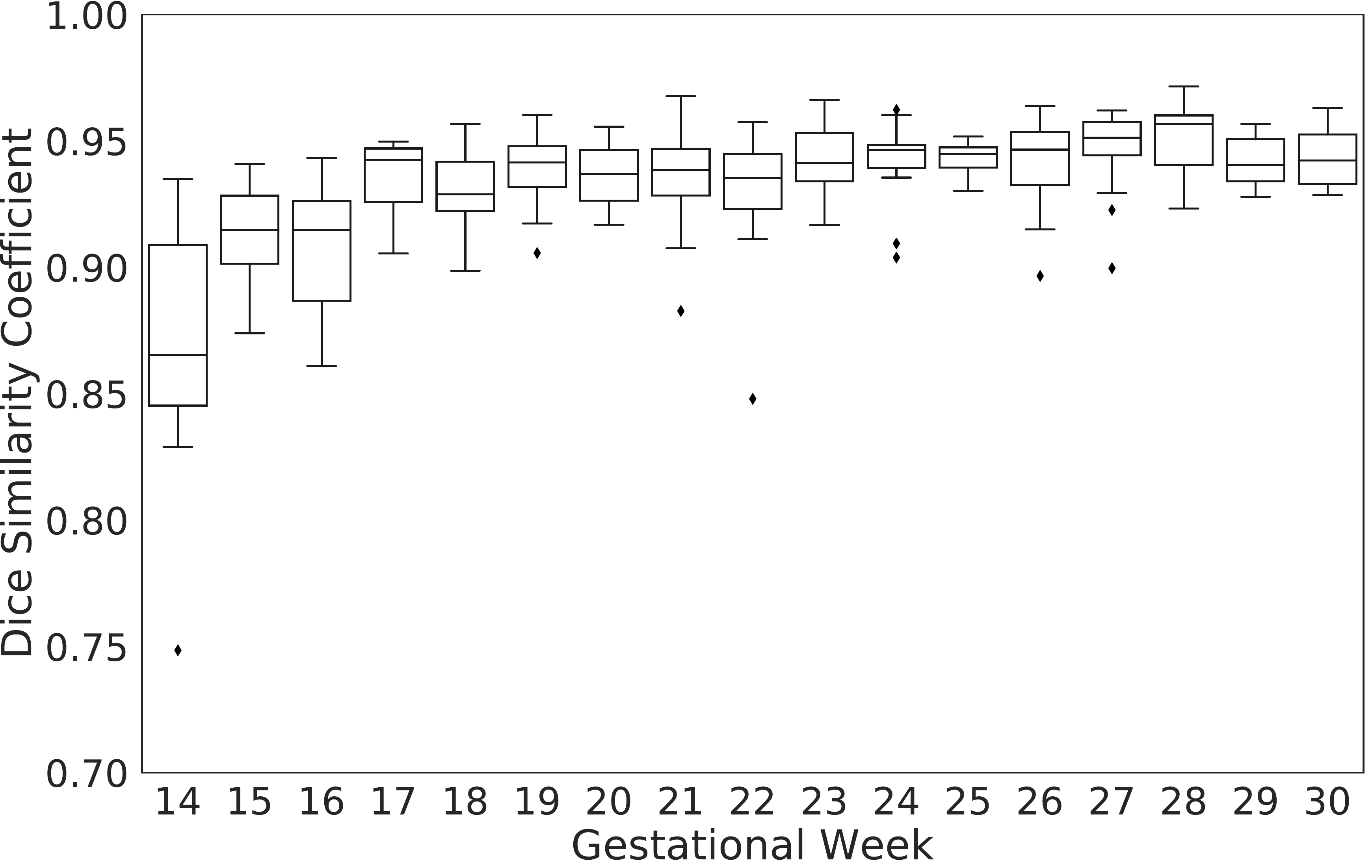} &
    \includegraphics[width=0.45\textwidth]{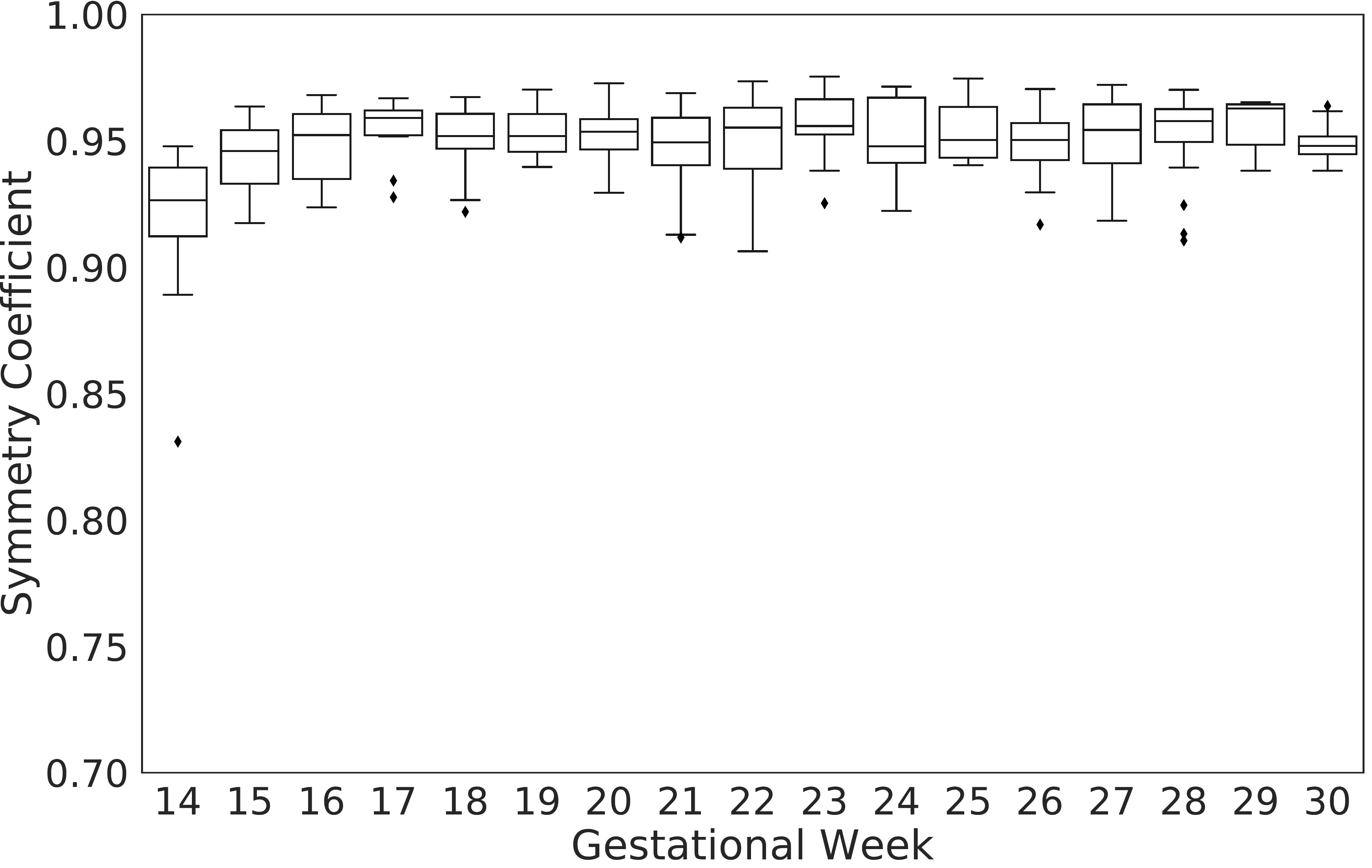}
\end{tabular}
  \caption{Network performance for each gestational week. ED is consistent throughout all ages. HD increases after week 21 most likely due to the different structural information obtained from US in comparison to the MRI-based annotation. DSC drops for weeks earlier than 21, as expected due to the annotations for these weeks being based on the atlas for 21 gestational weeks. SC shows no correlation with gestational age.\label{fig:gestationalresults}}
\end{figure}

The results describing the performance of the network for each gestational week are displayed in Fig. \ref{fig:gestationalresults}. The ED is consistently located between 1 mm and 3 mm throughout the data, with no observable correlation (r=0.14) with the gestational age of the fetus.
The HD is generally very consistent around 10 mm for weeks 14 to 22, with its value increasing for each week with a maximum of 15 mm at week 30. This is likely a result of the different structural information observed in fetal brain MRI and US as is described in Sec. \ref{sec:regionalperformance}. 
The DSC show a slight increase between weeks 14 and 18, with a very consistent behaviour with a weak correlation to gestational week (r=0.27). This is most likely caused by the atlas mask, since as mentioned in Sec. \ref{sec:data}, the atlas of week 19 was used for that gestational age range. Regardless, with the exception of week 14, all DSCs are in average consistently above 0.90.
Finally, the SC of the predicted volume is consistently high throughout the complete gestational period available in our data, with no significant correlation observed (r=0.19).

\subsection{Regional performance}\label{sec:regionalperformance}
\begin{figure}[htb!]
\centering
\begin{tabular}{m{0.9\linewidth}m{0.05\linewidth}}
  \begin{tabular}{ccc}
    \includegraphics[height=1.15cm]{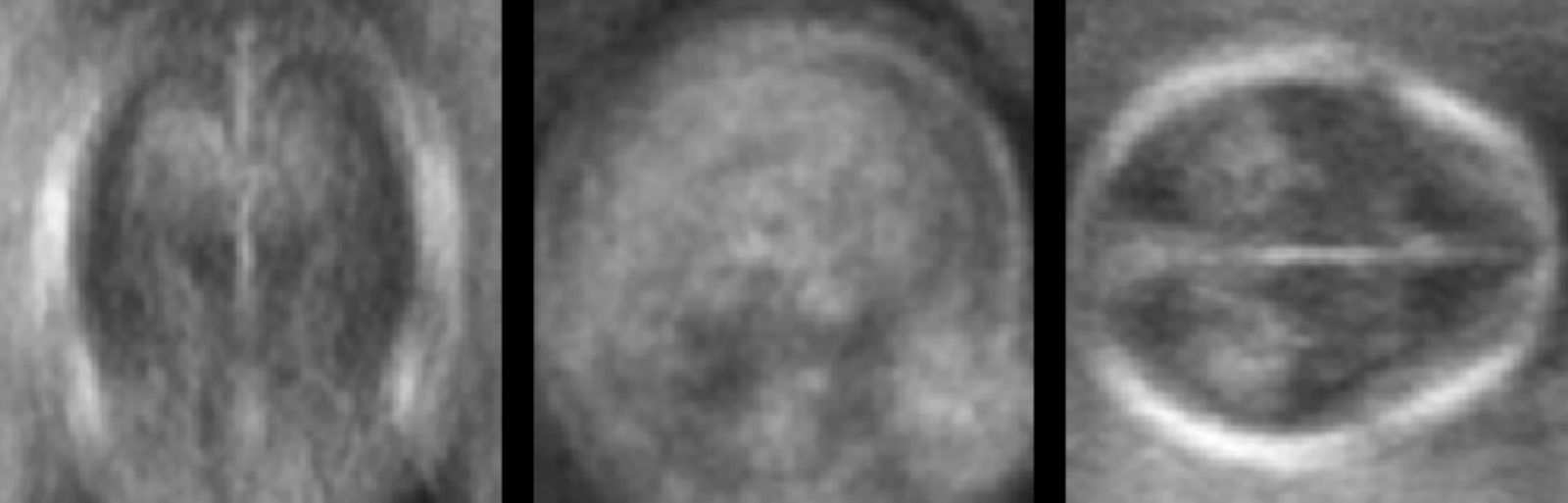} &
    \includegraphics[height=1.15cm]{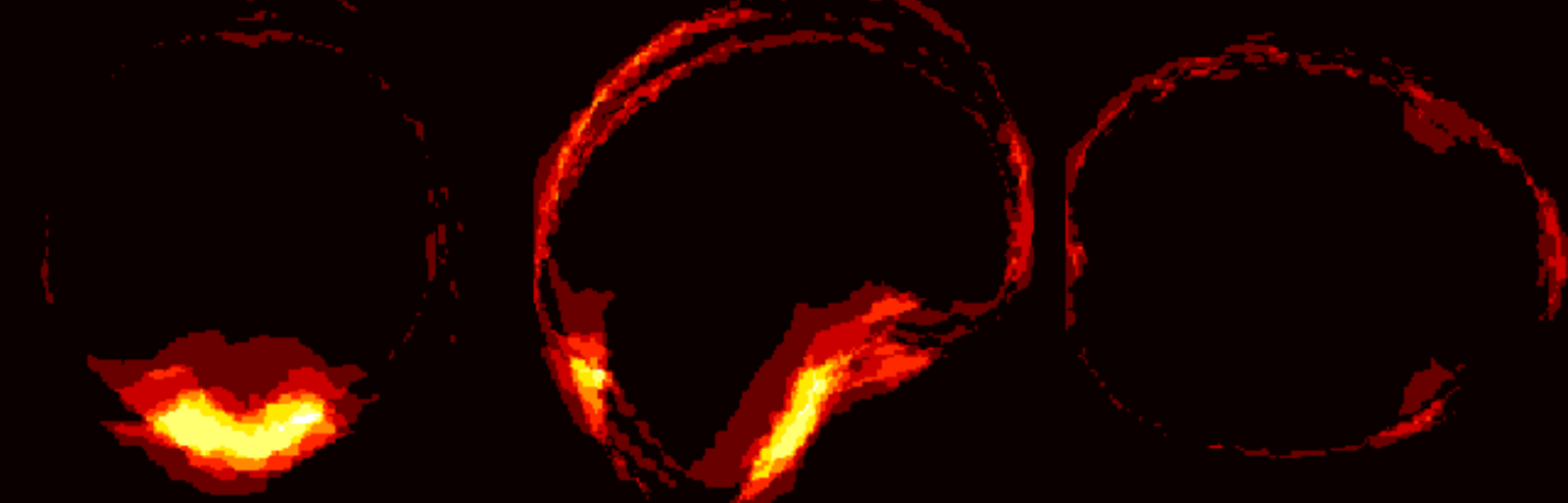} &
    \includegraphics[height=1.15cm]{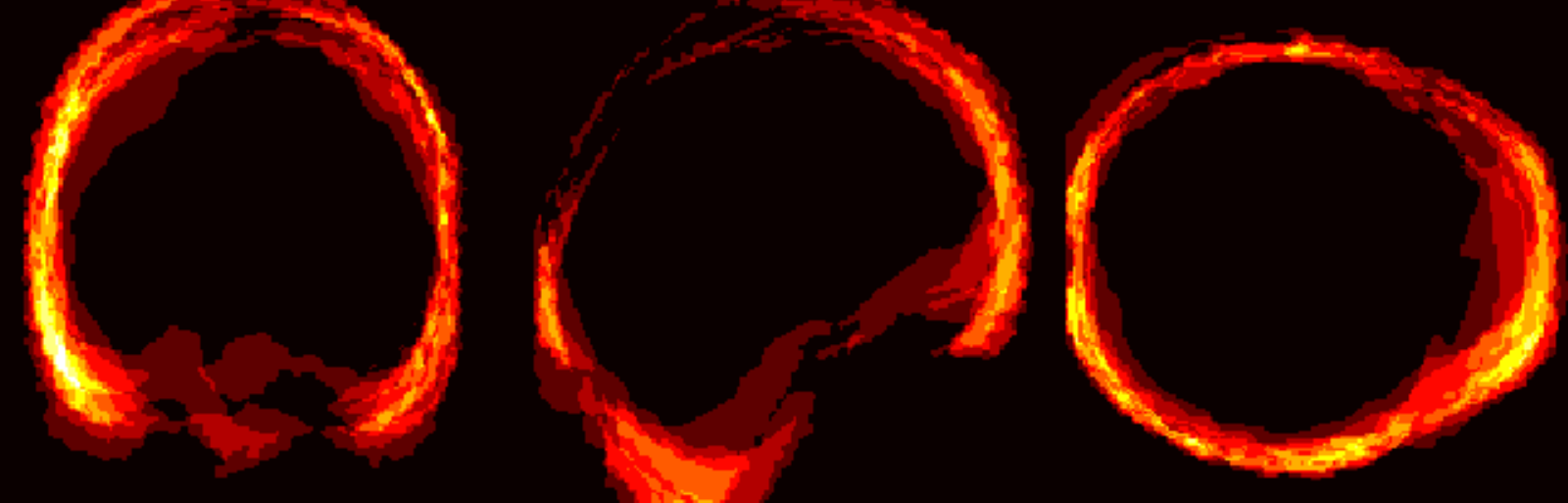}\\
    \includegraphics[height=1.15cm]{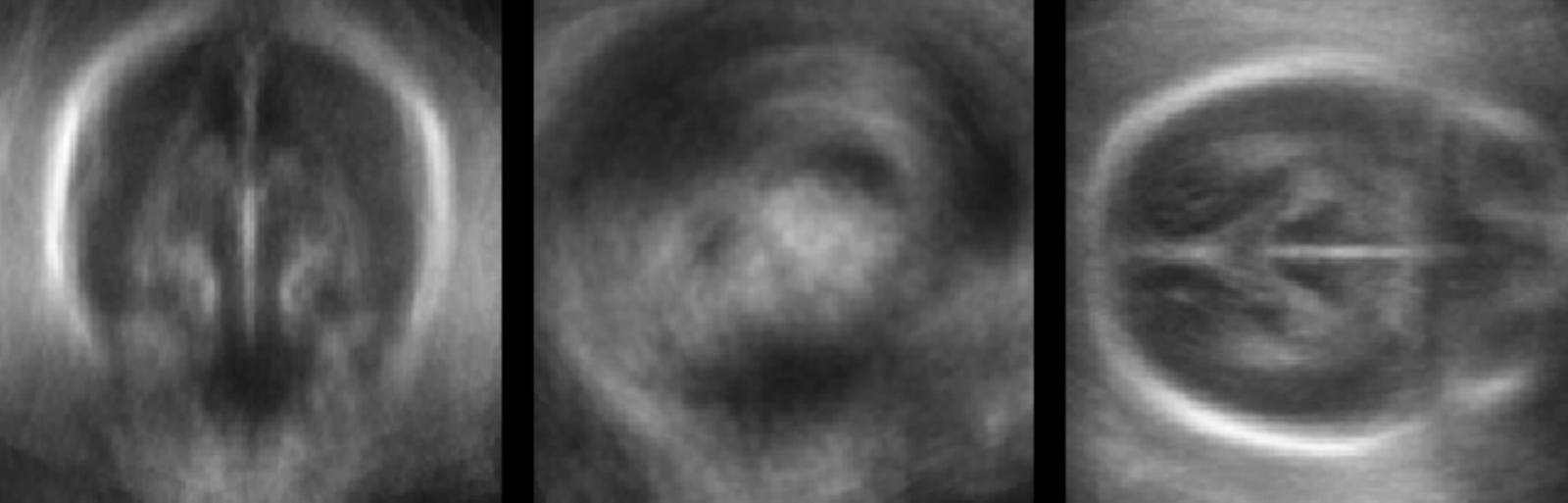} &
    \includegraphics[height=1.15cm]{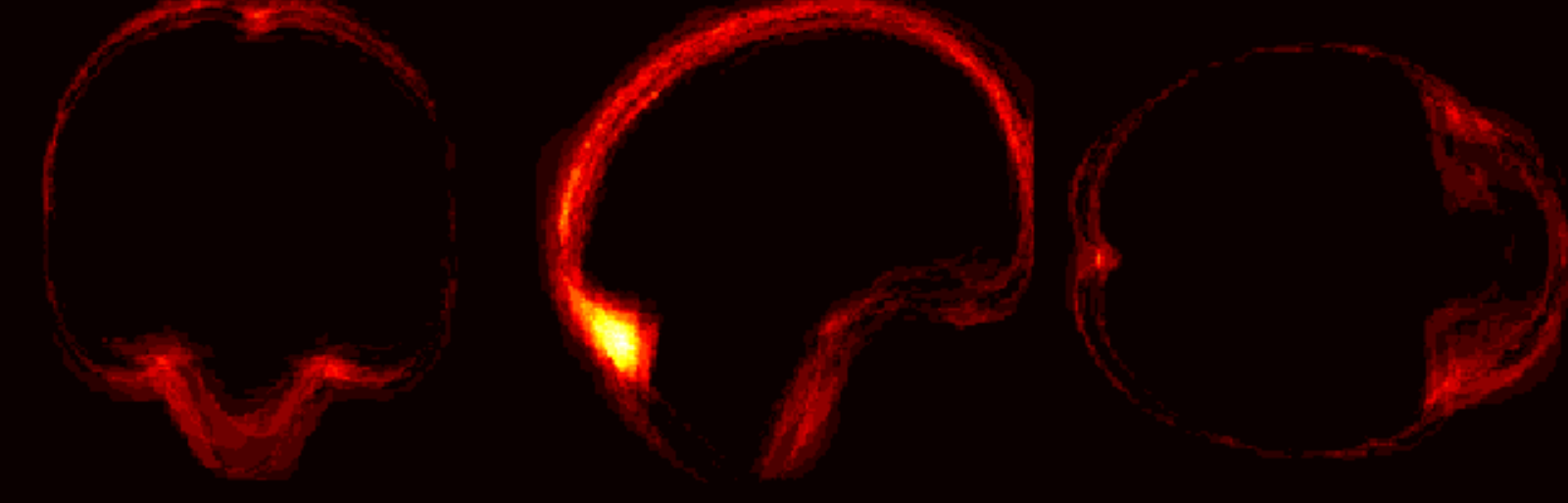} &
    \includegraphics[height=1.15cm]{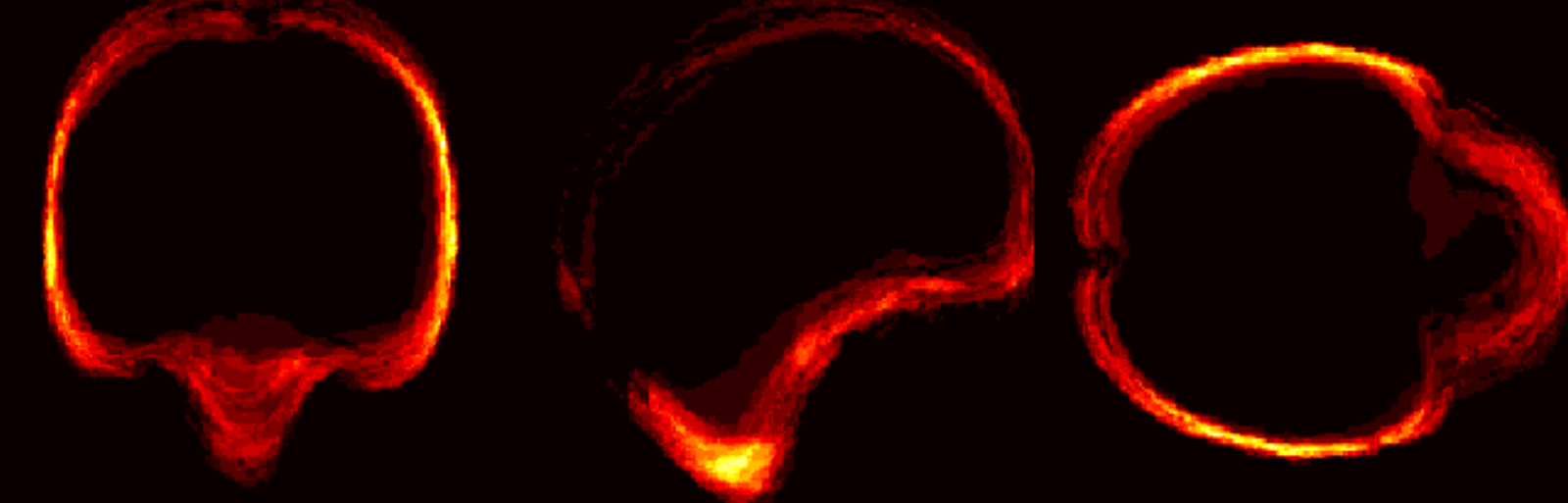}\\
    \includegraphics[height=1.15cm]{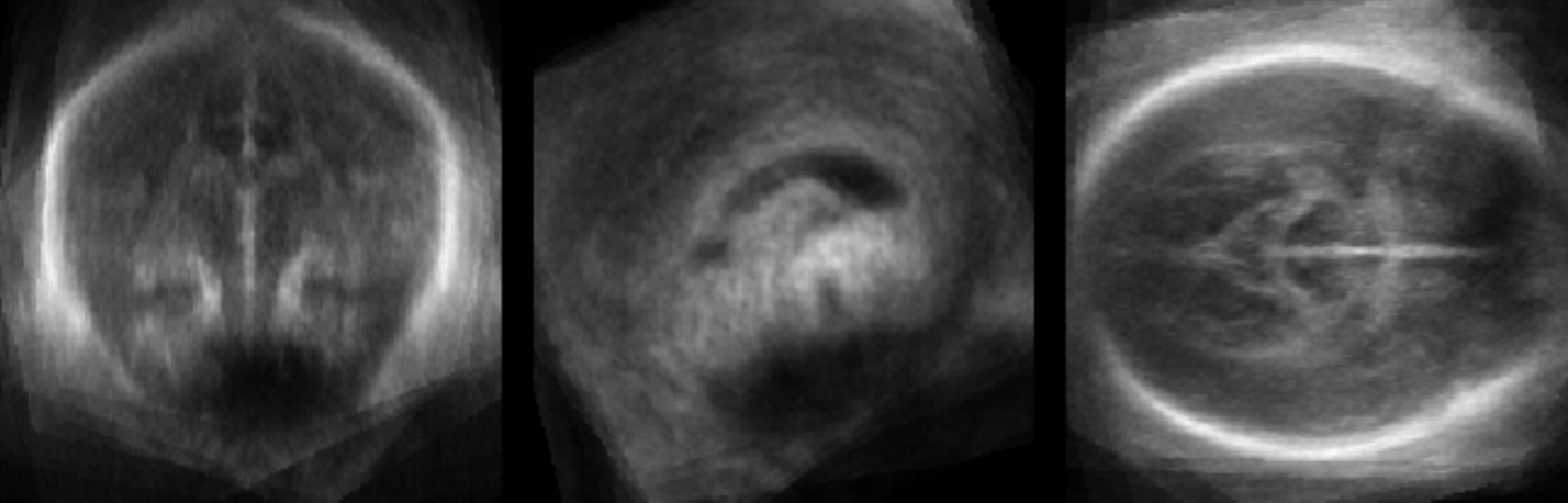} &
    \includegraphics[height=1.15cm]{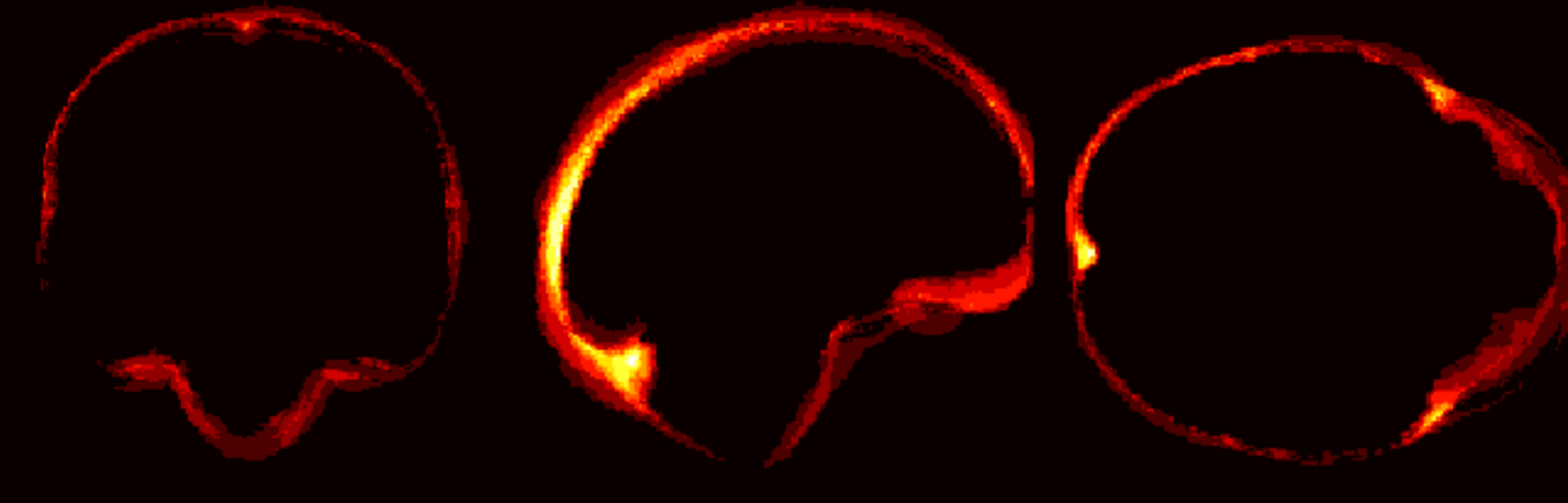} &
    \includegraphics[height=1.15cm]{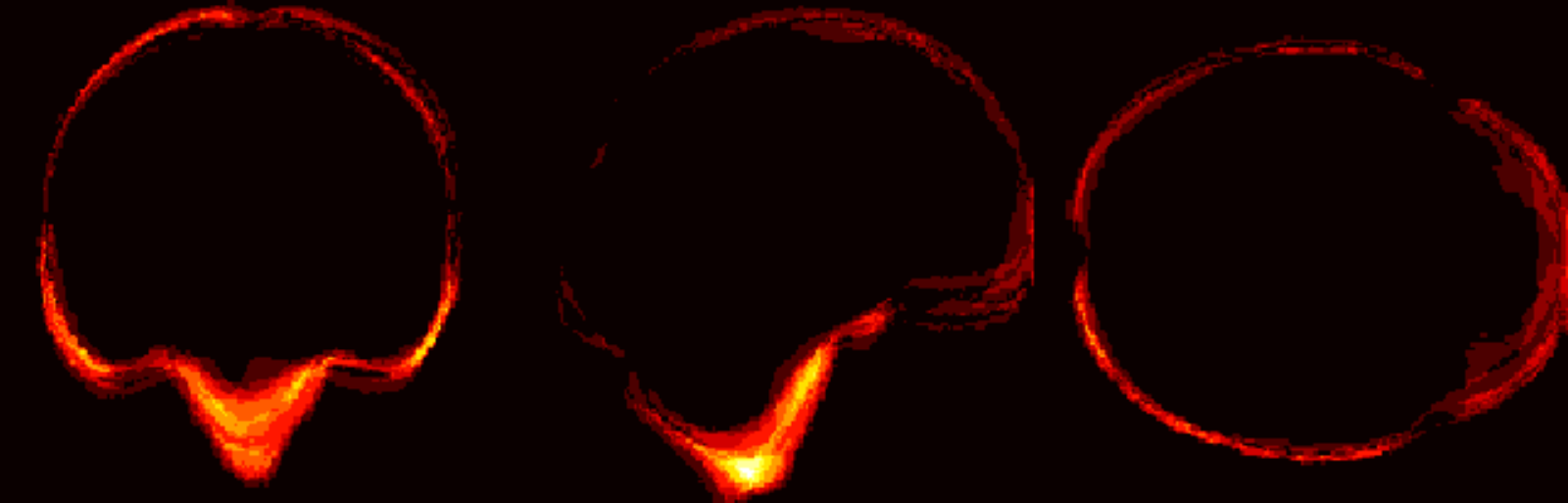} 
    \end{tabular}
	&
 \includegraphics[height=3.82cm]{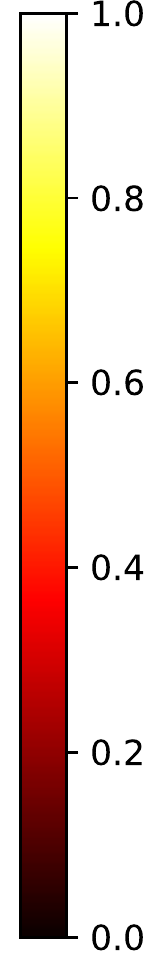}
    %Average Intensities & False Positives & False Negatives \\
\end{tabular}
  \caption{Regional performance comparison. False-positives (predicted voxels not present in the annotation) and false-negatives (non-predicted voxels present in the annotations) give a qualitative assessment of the performance of the predictions for particular regions of the brain. Top: 14 Weeks of gestation. Middle: 22 Weeks. Bottom: 30 Weeks. These gestational ages have been selected to represent the age range in our dataset and the results shown are the averages of all volumes tested in that gestational week. They have all been aligned and rigidly scaled to the same coordinate set. The False Positives and False Negatives values have been normalized to the number of volumes used in that gestational week. \label{fig:heatmaps}}
\end{figure}

The regional performance of the network is shown as a map of false-positives and false-negatives for different gestational ages in Fig. \ref{fig:heatmaps}. It can be observed that overall the network is very consistent throughout all regions of the brain with the exception of the brain stem. This is most likely due to this structural information not being visible in the US scan.
There is, however, one region of the brain that has worse regional performance for later gestational weeks: the space between the occipital cortex and the cerebellum. This is likely due to the fact that the annotations are based on the brain tissue and do not include the cerebrospinal fluid. As the brain develops, the separation between occipital cortex and cerebellum becomes more pronounced. However, since the US does not offer a good contrast between the tissue and the cerebrospinal fluid, this separation is not visible and therefore appears as a false positive.  

\subsection{Comparison with previous method}
\begin{figure}[!h]
\centering
  \begin{tabular}{ccccc}
   \includegraphics[height=5.2cm]{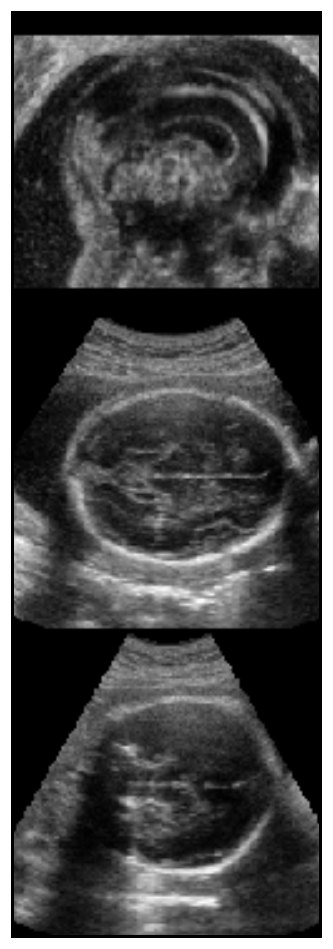} &
    \includegraphics[height=5.2cm]{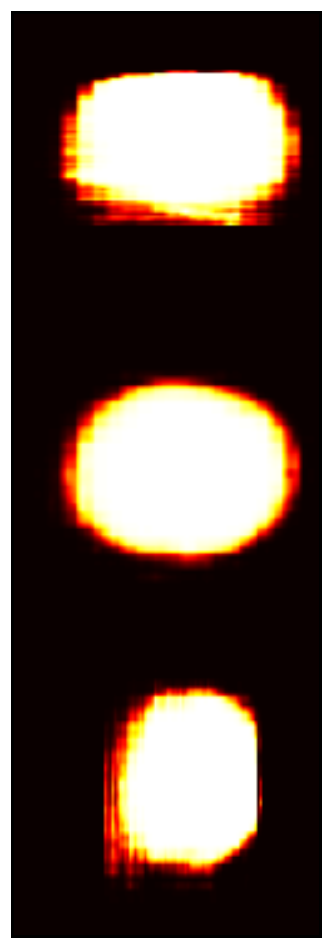} &
	\includegraphics[height=5.2cm]{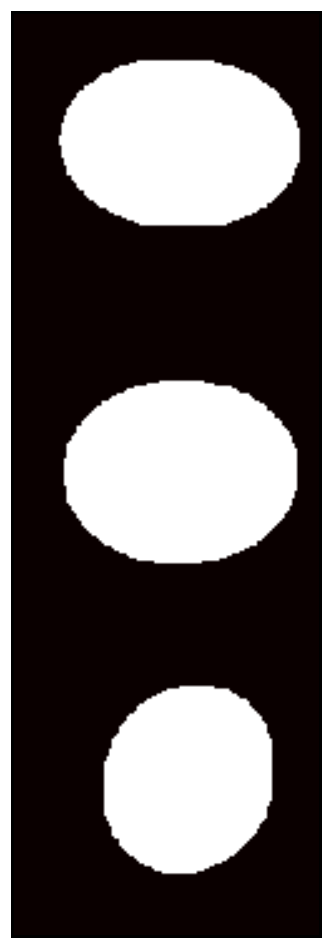} &
	\includegraphics[height=5.2cm]{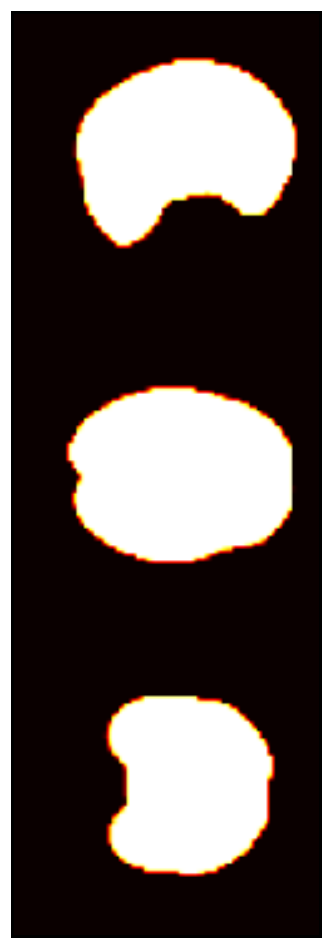} &
	\includegraphics[height=5.2cm]{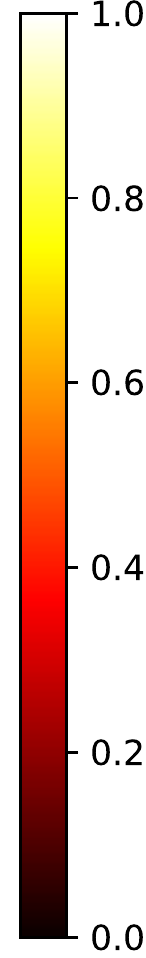}
	 \\
	Original & Namb. 01 & Namb. 02 & Our work& \\ 
\end{tabular}
  \caption{Comparison of our results with the method proposed by Namburete et. al., 2018~\cite{Namburete2018}. Original: the three mid-planes of the original volume. Namb. 01 and Namb. 02: Probability mask generated from the 2D predictions and the ellipsoid fiting respectively~\cite{Namburete2018}. Our work: The predictions obtained from our network. All columns but the first one share the same colormap.\label{fig:comparisonnamburete}}
\end{figure}

To quantitatively compare our network to the only other brain extraction method for fetal 3D US, the out test dataset was analysed with the method described in Namburete et al., 2018~\cite{Namburete2018}. The same experiments were performed and the results are shown in Table \ref{tab:comparisonnamburete}. Our network manages considerably better results throughout all comparisons. This is expected, since the other method relies on an approximation of the brain volume as ellipsoid, which does not accurately represent its shape. While an ellipsoid would be expected to have a high SC, its fitting to the probability mask results in an inaccurate alignment, which is reflected in the low SC of 0.74, compared to our network's 0.95. This can be clearly seen in Fig. \ref{fig:comparisonnamburete}, where a comparison of their method with ours is shown.

\begin{table}[!tb]
\caption{Network results against method from \cite{Namburete2018}. Our network shows consistently better results accross all evaluations. \label{tab:comparisonnamburete}}
\begin{tabular*}{\textwidth}{l@{\extracolsep{\fill}}lllll}
  \hline
Method & ED [mm]& HD [mm]& DSC & SC\\ \hline

	Our work& \SI{1.36 \pm 0.72}{} & \SI{9.05 \pm 3.56}{} & \SI{0.94 \pm 0.02}{} & \SI{0.95 \pm 0.02}{}\\
	Namburete et al., 2018~\cite{Namburete2018} & \SI{3.68 \pm 4.44}{} & \SI{15.12 \pm 5.24}{} & \SI{0.85 \pm 0.08}{} & \SI{0.74 \pm 0.06}{}\\

    \hline
\end{tabular*}
\end{table}

\section{Discussion and conclusion}
The method presented in this work was developed for automated brain extraction from fetal 3D US. For this, a 3D CNN was developed and optimized to predict where the brain is located without any need for pre- or post-processing, and its performance was analysed over a variety of conditions usually presented in 3D US, such as the variation in brain position, rotation, scale, as well as a variation in developmental age and therefore internal brain structures.

Our network provided consistent results throughout all experiments, managing to accurately locate the brain to a mean ED of 1.36 mm. As a comparison, this distance represents only 2\% of the mean occipitofrontal diameter of a brain of 22 weeks of gestation. The mean HD results of 9.05 mm is within what was expected due to the annotation masks being created from MRI data. The mean DSC of 0.94 shows that the predictions have a high level of overlap with the annotations and that our network manages consistent accurate predictions, which is also reflected on the mean SC of 0.95. The latter confirms that the network is not affected by the asymmetric structural information as a result of the beam-skull interactions.

In terms of pose variation, the network shows no correlation between the orientation of the brain and the accuracy of the prediction. The same consistency was shown throughout all gestational ages with the exception of HD for later gestational weeks and DSC for earlier gestational weeks. This is caused by two limitations of the proposed brain extraction tool: the annotation of the data coming from fetal MRI data, and the youngest available gestational age for these annotations being 19 weeks (see \ref{sec:data}). 
The fetal MRI data (and therefore our annotations) manage to separate the brain tissue and the cerebrospinal fluid, which is a distinction not visible in 3D US. This inherent difference between the two imaging modalities creates an annotation that does not accurately match our data around the edges of the mask. This is most pronounced in the space between the occipital cortex and the cerebellum, which becomes more pronounced as the gestational age increases, and is therefore reflected in the increased HD over time as well as in the regional performance analysis. The visibility of the brain stem in US is also not as good as in MRI, causing similar results. However, the fact that our network consistently extracts the brain including the cerebrospinal fluid is a good indication that our network has not been overfitted. 
As for the youngest gestational week available for annotations being week 19, it meant that our annotations for weeks 14-18 were less accurate. While the network results during this period is still very good (ED = 1.7 mm, HD = 7.14 mm, DSC = 0.92, SC = 0.95), it could be improved with a more accurate annotation.

The results of our network are significantly better than the ones obtained using the method from Namburete et al. 2018~\cite{Namburete2018}. This was expected due to that method using a 2D slice approach to the brain extraction predictions, and requiring an ellipsoid approximation of the brain volume, which does not represent the real structural characteristics of the organ.

In this work we have shown that our 3D CNN solution to fetal brain extraction from 3D US works accurately, reliably, and consistently, regardless of the large data variation inherent in this imaging modality.

\section*{Acknowledgement}
This work is supported by funding from the Engineering and Physical Sciences Research Council (EPSRC) and medical Research Council (MRC) [grant number EP/L016052/1].
\\
A. Namburete is grateful for support from the UK Royal Academy of Engineering under the Engineering for Development Research Fellowships scheme.
\\
B. W. Papie\.{z} acknowledges Rutherford Fund at Health Data Research UK.
\bibliographystyle{splncs04}

\begin{thebibliography}{8}

\bibitem{Kim2008}
Kim M.S., Jeanty P., Turner C., Benoit B.: Three-dimensional sonographic evaluations of embryonic brain development. J Ultrasound Med. 2008 Jan;27(1):119-24

\bibitem{Haratz2018}
Haratz K.K., Lerman-Sagie T.: Prenatal diagnosis of brainstem anomalies. Eur J Paediatr Neurol. 2018 Nov;22(6):1016-1026

\bibitem{Namburete2018-2}
Namburete A.I.L., Van Kampen R., Papageorghiou A.T., Papie\.{z} B.W.
Multi-channel groupwise registration to construct an ultrasound-specific fetal brain atlas. In: Melbourne A. et al. (eds) Data Driven Treatment Response Assessment and Preterm, Perinatal, and Paediatric Image Analysis. PIPPI 2018, DATRA 2018. Lecture Notes in Computer Science, vol 11076. Springer, Cham​

\bibitem{ISUOG2007}
International Society of Ultrasound in Obstetrics \& Gynecology Education Committee: Sonographic examination of the fetal central nervous system: guidelines for performing the 'basic examination' and the 'fetal neurosonogram'.Ultrasound Obstet Gynecol. 2007 Jan;29(1):109-16

\bibitem{Serag2016}
Serag A., Blesa M., Moore E. J., Pataky R., Sparrow S. A., Wilkinson A. G., Macnaught G., Semple S. I., Boardman, J. P. Accurate Learning with Few Atlases (ALFA): an algorithm for MRI neonatal brain extraction and comparison with 11 publicly available methods. Sci Rep. 2016; 6: 23470.


\bibitem{Ison2012}
Ison M., Dittrich E., Donner R., Kasprian G., Prayer D., Langs G.: Fully Automated Brain Extraction and Orientation in Raw Fetal MRI. Perinatal and Paediatric Imaging (PaPI 2012), MICCAI workshop

\bibitem{Keraudren2013}
Keraudren K., Kyriakopoulou V., Rutherford M., Hajnal J.V., Rueckert D.: Localisation of the Brain in Fetal MRI Using Bundled SIFT Features. Medical Image Computing and Computer-Assisted Intervention – MICCAI 2013. MICCAI 2013. 

\bibitem{Namburete2018}
Namburete A.I.L., Xie W., Yaquba M., Zisserman A., Noble J.A.: Fully-automated alignment of 3D fetal brain ultrasound to a canonical reference space using multi-task learning. Med Image Anal. 2018 May;46:1-14

\bibitem{Huang2018}
Huang R., Noble J.A., Namburete A.I.L.: Omni-Supervised Learning: Scaling Up to Large Unlabelled Medical Datasets. Medical Image Computing and Computer Assisted Intervention – MICCAI 2018. MICCAI 2018. 

\bibitem{Cicek2016}
Cicek O., Abdulkadir A., Lienkamp S.S., Brox T., Ronneberger O. 3D U-Net: learning dense volumetric segmentation from sparse annotation. Medical Image Computing and Computer-Assisted Intervention – MICCAI 2016

\bibitem{Papageorghiou2014}
Papageorghiou A.T., Ohuma E.O., Altman D.G., Todros T., Cheikh Ismail L., Lambert A., Jaffer Y.A., Bertino E., Gravett M.G., Purwar M., Noble J.A., Pang R., Victora C.G., Barros F.C., Carvalho M., Salomon L.J., Bhutta Z.A., Kennedy S.H., Villar J.; International Fetal and Newborn Growth Consortium for the 21st Century (INTERGROWTH-21st): International standards for fetal growth based on serial ultrasound measurements: the Fetal Growth Longitudinal Study of the INTERGROWTH-21st Project. Lancet. 2014 Sep 6;384(9946):869-79

\bibitem{Gholipour2017}
Gholipour A, Rollins CK, Velasco-Annis C, et al. A normative spatiotemporal MRI atlas of the fetal brain for automatic segmentation and analysis of early brain growth. Sci Rep. 2017 Mar;7(1):476

\bibitem{Moreira2011}
Moreira N.C., Teixeira J., Themudo R., Amini H., Axelsson O., Raininko R., Wikstrom J.: Measurements of the normal fetal brain at gestation weeks 17 to 23: a MRI study.Neuroradiology. 2011 Jan;53(1):43-8


%\bibitem{Toga2006}
%Toga A.W., Thompson P.M., Sowell E.R.: Mapping brain maturation. Trends Neurosci. 2006 Mar;29(3):148-59

%\bibitem{Namburete2015}
%Namburete A.I.L., Stebbing R.V., Kemp B., Yaqub M., Papageorghiou A.T., Noble J.A.: Learning-based prediction of gestational age from ultrasound images of the fetal brain. Medical Image Analysis 2015 Apr;21(1):72-86


%\bibitem{Namburete2017}
%Namburete A.I.L., Xie W., Noble J.A: Robust Regression of Brain Maturation from 3D Fetal Neurosonography Using CRNs. Fetal, Infant and Ophthalmic Medical Image Analysis. OMIA 2017, FIFI 2017.

%\bibitem{Benacerraf1998}
%Benacerraf B. R.: Ultrasound of Fetal Syndromes, Churchill Livingstone, 1998

%\bibitem{Conner2014}
%Conner S.N., Longman R.E., Cahill A.G.: The role of ultrasound in the diagnosis of fetal genetic syndromes. Best Pract Res Clin Obstet Gynaecol. 2014 Apr;28(3):417-28

%\bibitem{Kfir2009}
%Kfir M., Yevtushok L., Onishchenko S., Wertelecki W., Bakhireva L., Chambers C.D., Jones K.L., Hull A.D.: Can prenatal ultrasound detect the effects of in-utero alcohol exposure? A pilot study. Ultrasound Obstet Gynecol. 2009 Jun;33(6):683-9

%\bibitem{Kurtz1980}
%Kurtz A.B., Wapner R.J., Rubin C.S., Cole-Beuglet C, Ross R.D., Goldberg B.B.: Ultrasound criteria for in utero diagnosis of microcephaly.J Clin Ultrasound. 1980 Feb;8(1):11-6.

%\bibitem{Lee2010}
%Lee S., Walker S.P.: The role of ultrasound in the diagnosis and management of the growth restricted fetus. Australas J Ultrasound Med. 2010 Aug; 13(3): 31–36.  

%\bibitem{Monteagudo2009}
%Monteagudo A, Timor-Tritsch I.E.: Normal sonographic development of the central nervous system from the second trimester onwards using 2D, 3D and transvaginal sonography. Prenat Diagn. 2009 Apr;29(4):326-39

%\bibitem{Vinkesteijn2000}
%Vinkesteijn A.S., Mulder P.G., Wladimiroff J.W.: Fetal transverse cerebellar diameter measurements in normal and reduced fetal growth. Ultrasound Obstet Gynecol. 2000 Jan;15(1):47-51


%\bibitem{Yaqub2015}
%Yaqub M., Kelly B., Papageorghiou A.T., Noble J.A.: Guided random forests for identification of key fetal anatomy and image categorization in ultrasound scans. Medical Image Computing and Computer-Assisted Intervention – MICCAI 2015. MICCAI 2015. Bart\l{}omiej W. Papie\.{z}

%\bibitem{Qiu2017}
%Qiu W., Chen Y., Kishimoto J., de Ribaupierre S., Chiu B., Fenster A., Yuan J.: Automatic segmentation approach to extracting neonatal cerebral ventricles from 3D ultrasound images. Med Image Anal. 2017 Jan;35:181-191

%\bibitem{Namburete2014}
%Namburete A.I.L., Yaqub M., Kemp B., Papageorghiou A.T., Noble J.A.: Predicting fetal neurodevelopmental age from ultrasound images. Med Image Comput Comput Assist Interv. 2014;17(Pt 2):260-7

%\bibitem{Pistorius2010}
%Pistorius L.R., Stoutenbeek P., Groenendaal F., de Vries L., Manten G., Mulder E., Visser G.: Grade and symmetry of normal fetal cortical development: a longitudinal two- and three-dimensional ultrasound study. Ultrasound Obstet Gynecol. 2010 Dec;36(6):700-8


%\bibitem{Kleesiek2016}
%Kleesiek J., Urban G., Hubert A., Schwarz D., Maier-Hein K., Bendszus M., Biller A.: Deep MRI brain extraction: A 3D convolutional neural network for skull stripping. Neuroimage. 2016 Apr 1;129:460-469


%\bibitem{Row2018}
%Roy S., Knutsen A., Korotcov A., Bosomtwi A., Dardzinski B., Butman J.A., Pham D.L.: A deep learning framework for brain extraction in humans and animals with traumatic brain injury . 2018 IEEE 15th International Symposium on Biomedical Imaging (ISBI 2018) 


%\bibitem{Yaqub2016}
%Yaqub M., Rueda S., Kopuri A., Melo P., Papageorghiou A.T., Sullivan P.B., McCormick K., Noble J.A.: Plane Localization in 3-D Fetal Neurosonography for Longitudinal Analysis of the Developing Brain. IEEE J Biomed Health Inform. 2016 Jul;20(4):1120-8

%\bibitem{Toi2004}
%Toi A., Lister W.S., Fong K.W.: How early are fetal cerebral sulci visible at prenatal ultrasound and what is the normal pattern of early fetal sulcal development? Ultrasound Obstet Gynecol. 2004 Dec;24(7):706-15
%
%\bibitem{Timor-Tritsch2000}
%Timor-Tritsch I.E., Monteagudo A., Mayberry P.: Three-dimensional ultrasound evaluation of the fetal brain: the three horn view. Ultrasound Obstet Gynecol. 2000 Sep;16(4):302-6



%\bibitem{Malinger2004}
%Malinger G., Ben-Sira L., Lev D., Ben-Aroya Z., Kidron D., Lerman-Sagie T. Fetal brain imaging: a comparison between magnetic resonance imaging and dedicated neurosonography. Ultrasound Obstet Gynecol. 2004 Apr;23(4):333-40



\end{thebibliography}
%

\end{document}